\documentclass[12pt]{JHEP3}

\usepackage{amssymb, amsmath, amsopn, amsthm}
\usepackage{epsfig}
\usepackage{graphics}

\def\makeatletter{\catcode`\@=11}
\makeatletter
\def\mathbox#1{\hbox{$\m@th#1$}}%
\def\math@ccstyles#1#2#3#4#5#6#7{{\leavevmode
      \setbox0\mathbox{#6#7}%
      \setbox2\mathbox{#4#5}%
      \dimen@ #3%
      \baselineskip\z@\lineskiplimit#1\lineskip\z@
      \vbox{\ialign{##\crcr
             \hfil \kern #2\box2 \hfil\crcr
             \noalign{\kern\dimen@}%
             \hfil\box0\hfil\crcr}}}}
\def\mathaccstyles{\math@ccstyles\maxdimen}
\def\maththroughstyles{\math@ccstyles{-\maxdimen}}
\def\unity%
 {\maththroughstyles{.45\ht0}\z@\displaystyle {\mathchar"006C}\displaystyle 1}

\title{Holographic Fermi and Non-Fermi Liquids with Transitions in Dilaton Gravity}

\author{Norihiro Iizuka$^{1}$,~Nilay Kundu$^{2}$,~Prithvi Narayan$^{2}$~and Sandip P. Trivedi$^{2}$ 

  ~\\
  $^1$Theory Division, CERN, CH-1211 Geneva 23, Switzerland \\
norihiro.iizuka@cern.ch \\

$^2$Tata Institute for Fundamental Research \\
Homi Bhabha Road, Mumbai 400005, India\\
prithvi.narayan@gmail.com, nilay.tifr@gmail.com, trivedi.sp@gmail.com

\vspace{0.1cm}

}

\abstract{ We study the  two-point function for fermionic operators 
 in a class of strongly coupled systems using the gauge-gravity correspondence. The 
gravity description includes a  gauge field and a  dilaton which determines  the gauge coupling
and the potential energy.  
Extremal black brane solutions in this system typically have vanishing entropy. 
By analyzing a charged fermion in these extremal black brane backgrounds we calculate the 
two-point function of the corresponding boundary fermionic operator. We find that in some region of parameter space it is of 
Fermi liquid type. 
Outside this region no well-defined quasi-particles exist,  with the excitations acquiring a non-vanishing
width  at zero frequency.
At the transition,  the two-point function  can exhibit non-Fermi liquid  behaviour.

}

\preprint{CERN-PH-TH-2011-101 \\ TIFR/TH/11-20}

\newcommand{\ra}{\rightarrow}
\newcommand{\be}{\begin{equation}}
\newcommand{\ee}{\end{equation}}
\newcommand{\bc}{\begin{center}}
\newcommand{\ec}{\end{center}}
\newcommand{\ba}{\begin{align}}

\newcommand{\tz}{{\tilde z}}

\begin{document}

\tableofcontents




\section{Introduction}

The  Gauge/Gravity correspondence \cite{Maldacena:1997re}, \cite{Gubser:1998bc}, \cite{Witten:1998qj} provides us with a new tool to study strongly coupled field theories. 
It is  worth exploring whether  insights of relevance to condensed matter physics can be gained  using this tool. 
One set of questions which have proved difficult to analyze using conventional techniques is the behaviour of fermions in strongly coupled 
systems in the presences of a chemical potential. There has been considerable  activity exploring  this issue on the 
the  gravitational side recently and some interesting lessons have been learnt from such studies. 
This question is  particularly interesting  in view of considerable evidence now  for non-Fermi liquid behaviour 
in condensed matter systems, e.g., in High $T_c$  materials and in heavy 
fermion systems close to quantum phase transitions.

Extremal black branes, which are at non-zero chemical potential and typically at zero temperature, 
 are of particular interest in the gravity description in exploring this question.  Some of the early studies 
 have focussed on analyzing the behavior of fermionic fields in  extremal Reissner Nordstrom (eRN) Black brane backgrounds \cite{SSL},
\cite{MIT}, \cite{CSZ}, \cite{MIT2}, \cite{MIT3}, \cite{MIT4}. 
While these black branes have the virtue of being simple and explicit they suffer from an important unphysical feature, namely, 
their entropy does not vanish  despite their vanishing temperature.  
 Instead, the  entropy of these extremal solutions 
 scales with the appropriate power of the chemical potential and  increases as the chemical potential increases. 
It is widely believed that this big  violation of the third law of thermodynamics is an artifact of the large $N$ limit, and in the absence of supersymmetry or say the  infinite number of symmetries  in 1+1 dim. CFT's, this degeneracy should be lifted once finite $N$ 
corrections are included 
\footnote{
For some discussion of related issues in extremal black holes see \cite{DST}.}.
 In this context it is also relevant to note  that 
quite often in string constructions extremal RN black branes have been found to be  unstable, 
for example due to the presence of light charged scalars, \cite{Gubser1}. 

In this paper we will consider $3+1$ gravity systems with a  holographic field theory dual which is $2+1$ dimensional.
The reservations discussed above for extreme RN black branes  make it worth looking for  other  gravity systems where  the extremal black branes are different  so that their entropy in particular vanishes at extremality.  
Such a class of systems was explored in \cite{Gubser:2009qt}, \cite{GKPT}, \cite{GIKPTW}. 
The key new ingredient was to include a dilaton which allows the 
gauge coupling of the Maxwell field to vary. It was found that as a  result the black branes have  zero entropy at extremality 
  \footnote{More precisely to ensure that higher derivative corrections are small one should introduce a small temperature.
One then finds that the entropy density vanishes as a positive power of the temperature.}.  
The dilatonic systems were further generalized  by \cite{Kiritsis}, see also, \cite{MGK},
 with both the gauge coupling and the potential energy now 
depending on the  dilaton. Extremal black branes were often found to posses zero entropy in such systems as well.

These dilatonic systems,  in particular their extremal black brane solution, are therefore  a promising starting point for exploring 
questions related to the behaviour of fermionic fields. The behaviour of a  bulk fermion in 
an  extremal black brane solution of the type studied in  \cite{GKPT}, \cite{GIKPTW},  was analyzed  in \cite{FP}, 
\cite{PT}.  It was found that the two-point function of the corresponding  fermionic operator in the boundary theory was 
qualitatively quite different from the non-Fermi liquid behaviour found in the eRN case and much more akin to a Fermi-liquid. 
The two-point function showed that there is a sharp Fermi surface in the system 
 with   well-defined quasi-particles excitations  which have  a linear, i.e., relativistic,   dispersion relation 
at small frequency,
and  a width
which has an essential singularity at vanishing frequency and  which  is therefore  very narrow at small frequency \footnote{
  For a  Fermi liquid the width is $O(\omega^2)$,
which is much broader. It could easily be that 
additional interactions, e.g., of $4$-Fermi type, which are suppressed in the large $N$ limit, when incorporated  can 
broaden out this width to the $\omega^2$ behaviour of Fermi liquid theory. 
 Keeping 
this in mind we will refer to such behaviour   as  being of Fermi-liquid type  below. It is also worth mentioning that there are
many  additional gapless excitations in the system which contribute to the specific heat and the conductivity. For this reason such a phase
is described as a fractionalised Fermi-Liquid (FL*) phase rather than a Fermi Liquid phase in \cite{Sachdev1}, \cite{Sachdev2},
\cite{Sachdev3}.}.  

In this paper we will analyze the behaviour of a charged  bulk fermion for the more general class of extremal dilaton systems studied in 
\cite{Kiritsis} and use it to calculate the two-point function of the corresponding fermionic operator in the boundary. 
We find that there is a wide range of behaviours that the fermion two-point function exhibits. 
One parameter in particular determines this behaviour, it is denoted by $\beta+\gamma$ below (for a definition of 
 $\beta,\gamma$ in terms of the parameters
appearing in the Lagrangian eq.(\ref{action}), see  eq.(\ref{case1}), (\ref{case11})). For $\beta+\gamma>1$ 
 one gets Fermi-liquid behaviour. At $\beta+\gamma=1$ there is a transition. For $\beta+\gamma<1$ there are no well-defined 
quasi-particle excitations since they 
 acquire
a big width which is non- vanishing as $\omega\rightarrow 0$ \footnote{For a more precise description see \S3.5 and also \S4.}. 
The behaviour at the transition, when $\beta+\gamma=1$, is also quite interesting.  The geometries which correspond to this case include
both extreme RN type solutions and other backgrounds where the entropy vanishes. These additional backgrounds, we find, 
 also give rise to non-Fermi liquid behaviour of a type very similar to that seen in the extreme RN case first.

From the field theory point of view, the systems we analyze can be thought of as essentially free  fermions  coupled to a 
fermionic operator of a strongly interacting sector \cite{FP}. The near-horizon gravity solution provides a dual  description
of  the strongly coupled sector. 
By varying the parameters $\beta, \gamma$ we explore different kinds of strongly coupled sectors and the resulting change in 
the behaviour of the fermionic two-point function. Our central result is that the class of strongly coupled sectors which are described by 
our gravity backgrounds can give rise to the different types of behaviour mentioned above and to transitions among these kinds of behaviours. 
For example, at the risk of belaboring this point, 
our results  show that Non-Fermi liquid behaviour of the type found first in \cite{SSL} - \cite{MIT4} is more common and can occur without the large entropy of the 
eRN case. It also shows that transitions can occur across which  well-defined quasi-particle excitations acquire a big width and cease to exist. 

On general grounds we expect to be able to model only strongly coupled systems in the large $N$ limit using a 
classical gravity description. 
  This is a central  limitation of our analysis. As a result  the fermions we are studying  are only a small subsector 
of a much bigger system with many degrees of freedom.  It turns out that while the fermionic two-point function
undergoes dramatic changes as  the parameter $\beta+\gamma$ is varied, as was mentioned above, 
 the geometry and other background fields change smoothly, signalling that  
most of the degrees of freedom of  the large $N$ ``heat bath'' in fact  do not change their behaviour is a significant way. 
As a result, one finds that the thermodynamics and transport properties like electrical conductivity also 
 do not change significantly;  in particular
the  qualitatively big changes in the fermionic two-point function do not correspond to  phase transitions. 
Once one goes beyond the large $N$ limit one expects that the significant  changes in the behaviour of the fermions, 
should they continue to occur, would also be   accompanied by significant changes in thermodynamics and transport. 
A preliminary indication of this is provided by $1/N$ corrections to the 
electrical conductivity which is sensitive to the change in the Fermion two-point function and 
therefore to a change in its properties, \cite{MIT3}, \cite{MIT4}.  The results of this paper   showing  
 that   behaviours other than of Fermi liquid type can arise in a fairly robust way
  may be taken as preliminary evidence that such behaviour is fairly generic in strongly coupled field theories
and   could   occur   beyond the large $N$ limit as well. 

Before proceeding it is worth commenting on some of the related literature. 
For a   general discussion  about phase  transitions where the Fermi surface  disappears and  non-Fermi liquid behaviour  can arise
see \cite{Senthil}.  
A  system with fermions living on probe branes with   examples of  
  Non-Fermi liquid phases and   transitions  due to the excitations getting gapped was found in \cite{KKY}. 
Some discussion of  the Holographic description of a Fermi liquid can be found in \cite{KP}. Progress towards constructing  an 
  holographic description of the strange metal phase  can be found in  \cite{strange}. 
Recent progress in understanding  the  Holographic  non-Fermi liquid phases often found in gravity systems  
 in terms of fractionalised Fermi liquids and related ideas is  
 contained in \cite{Sachdev1}, \cite{Sachdev2},
\cite{Sachdev3}.

This paper is organized as follows. We begin by reviewing the dilaton system of interest and discuss the near-horizon geometry of
extremal and near-extremal black branes in this system, along with some aspects of 
 their thermodynamics and transport  in \S2. The fermionic two-point function,  for various ranges of parameters, is discussed in \S3.
\S4 contains a summary of main results and conclusions. Appendix A contains a discussion of additional extremal solutions,
Appendices B - D contain additional material useful for the fermion two-point function calculation and Appendix  E 
contains the calculation for the scalar two-point function. In Appendix F we obtain a numerical solution interpolating between the near horizon geometry and $AdS_4$.

\section{The Dilaton Gravity System }

The system we consider  
consists of gravity, a  $U(1)$ gauge field, and a  scalar, $\phi$, which we call the dilaton, with   action, 
\be
\label{genact}
S = \int d^4\!x \sqrt{-g} \left\lbrace R - 2 (\nabla \phi)^2 -f(\phi) F_{\mu \nu} F^{\mu \nu} - V(\phi) \right\rbrace.
\ee
Note that for simplicity we have taken the kinetic energy term of the  dilaton to be canonical.
  This restriction can be easily relaxed although we will not do so here. 

The gauge coupling 
$g^2
\equiv (f(\phi))^{-1}$ and the potential $V(\phi)$ are both a function of the dilaton. 

We will be particularly interested in solutions
 where the dilaton has a run-away type of behaviour near the horizon of an extremal black brane. 
Such run-away behaviour can result in the entropy of the extremal brane vanishing \cite{GKPT}. 
Also, we will be  mainly  concerned   with the low-temperature or  low frequency (compared to the chemical potential) 
response of the system. On general grounds one expects that this response will  be   determined by the near-horizon geometry. 
Thus for our purposes we will  mainly be interested in the behaviour of $f(\phi)$ and $V(\phi)$ when the dilaton has evolved sufficiently far 
 along the run-away direction. We will take this behaviour to be of exponential type,  
\be
\label{dildepf}
f(\phi)=e^{2\alpha \phi},
\ee
\be
\label{dildeppot}
V(\phi)=V_0 e^{2\delta \phi}. 
\ee
The parameters $\alpha, \delta$ thus characterize the run-away behaviour which occurs for $\phi =\pm \infty$.
These parameters will  repeatedly enter the discussion below. 
Substituting in eq.(\ref{genact}) then gives the action, 
\begin{equation}
\label{action}
S = \int d^4\!x \sqrt{-g} \left\lbrace R - 2 (\nabla \phi)^2 -e^{2 \alpha \phi} F_{\mu \nu} F^{\mu \nu} - V_0 e^{2 \delta \phi}\right\rbrace
\end{equation}

It is worth noting before we proceed that the full dependence of $f(\phi), V(\phi)$ away from the run-away region
 can be very different from these exponential forms. In fact, to obtain a solution which is asymptotically $AdS_4$ space, 
 the potential $V(\phi)$ will need to have an extremum at a negative value of the cosmological constant and the dilaton will have to 
asymptote to this extremum far away from the horizon.  However, these features of $f(\phi), V(\phi)$, 
  and the corresponding features of the geometry,
will not be very significant for determining the low-energy behaviour which will arise essentially from the near-horizon region. 
In field theory terminology these features correspond to UV data  which  is irrelevant for   IR physics. The action
eq.(\ref{action}) therefore determines only the IR physics of the field theory.  
In the analysis below we will also take $V_0$ appearing in eq.(\ref{dildeppot}) to satisfy the condition \footnote{Since a negative cosmological constant is easier to obtain in string/M theory this choice for the sign of $V_0$
 might be also easier to obtain in a string/M construction.}
, 
\be
\label{condv0}
V_0<0. 
\ee
 
In this paper we will be interested in electrically charge black branes. Using the expected symmetries of the solution
(translations and rotation in the $x,y$ directions and time independence), the metric can be chosen to be of the form, 
\be
\label{ansatz1}
 ds^2 = - a(r)^2 dt^2 + { dr^2 \over a(r)^2 } + b(r)^2 (dx^2 + dy^2)
\ee
The horizon of the extremal  black brane will be taken to   lie at $r=0$. 
The gauge field equation of motion gives, 
\be
\label{gf}
 F = {Q_e \over f(\phi) b^2 }  dt \wedge dr.
\ee

The remaining equations of motion can be conveniently expressed in terms of an effective potential as \cite{GIJT}
\begin{equation}\label{Veff}
V_{\text{eff}} = {1 \over b^2} \left( e^{-2 \alpha \phi} Q_e^2  \right) + {b^2 V_0  \over 2} e^{2 \delta \phi},
\end{equation}
and are given by, 
\begin{align}
 \label{em1}
 (a^2 b^2)'' = & -2 V_0 e^{2 \delta \phi} b^2 \\
 \label{em2}
 {b'' \over b} = & -  \phi'^2 \\
 \label{em3}
 (a^2 b^2  \phi')'   = &   {1 \over 2} \partial_\phi V_{\text{eff}} \\
\label{em4}
 a^2 b'^2 + {1 \over 2} {a^{2}}' {b^{2}}' = &  a^2 b^2 \phi'^2 - V_{\text{eff}}.
\end{align}

\subsection{The Solutions}
In this subsection we will construct the near-horizon geometry for a class of extremal black brane solutions to 
these equations. 
Consider an  
ansatz \footnote{Actually the only assumption in this ansatz is that $b$ has a  power law dependence on $r$. Given this fact, 
eq. (\ref{em2})  implies that 
 $\phi \propto \log r$,  and eq.(\ref{em1}) implies (except for some special cases)  that $a(r)$ is also a  power law.}
\begin{equation}\label{ansatz2}
a = C_a r^{\gamma} \hspace{10mm} b= r^\beta \hspace{10mm} \phi = k \log{r}
\end{equation}
Note that a  multiplicative  constant  in   $b$ can be set to unity by rescaling $x,y$, 
and an additive constant in $\phi$, while subdominant at small $r$ can be absorbed into $V_0$ and $Q$. 
With this ansatz, eq.(\ref{em2}) gives,
\begin{equation}
\label{condk}
k^2 = \beta (1 - \beta)
\end{equation}
Eq.(\ref{em1}) gives, 
\begin{eqnarray}
C_a^2 (1 + \beta + k \delta) (1 + 2 \beta + 2 k \delta) &=& - V_0 \label{cond1ca}\\
\gamma -1 &=& \delta k \label{condgamma}. 
\end{eqnarray}
Finally, eq.(\ref{em3}), eq.(\ref{em4}) give,  on using $\gamma=1+\delta k$ and eq.(\ref{condk}),
\begin{equation}
\label{rel3}
\left(C_a^2 k ( 1+2 \beta + 2 k \delta) -{\delta \over 2} V_0\right) r^{4\beta-2k(\alpha-\delta)}  =  - \alpha Q_e^2 r^{-4 \alpha k} 
\end{equation}
and, 
\begin{equation}
\label{rel4}
\left(  C_a^2 \beta (1 + 2 \beta + 2 k \delta ) + {1 \over 2} V_0 \right) r^{4 \beta - 2k(\alpha - \delta)}  =    -  Q_e^2 r^{-4 \alpha k} 
\end{equation}
These last two equations can be met if  the power of $r$ and the coefficients match on both sides of each equation. 
This gives,
\begin{eqnarray}
4\beta & =& -2k(\alpha+\delta) \label{condbeta}\\
C_a^2  k (1 + 2 \beta + 2 k \delta) -{\delta \over 2} V_0&=& - \alpha Q_e^2 \label{cond2ca} \\
C_a^2 \beta (1 + 2 \beta + 2 k \delta ) + {1 \over 2} V_0&=& - Q_e^2 \label{cond3ca}
\end{eqnarray}

Solving,  eq.(\ref{condk}), eq.(\ref{cond1ca}), eq.(\ref{condgamma}),
 eq.(\ref{condbeta}), eq.(\ref{cond2ca}), eq.(\ref{cond3ca}),
 gives
\begin{align}\label{case1}
\beta = { (\alpha+\delta)^2 \over 4 + (\alpha+\delta)^2}   \hspace{10mm} 
\gamma = 1 -{ 2 \delta (\alpha+\delta) \over 4 + (\alpha+\delta)^2}   &\hspace{10mm} 
 k = - { 2 (\alpha+\delta) \over 4 + (\alpha+\delta)^2}  
\\
\label{case11} C_a^2  = -V_0 {\left( 4 + (\alpha+\delta)^2 \right)^2 \over 2 \left(2 + \alpha (\alpha+\delta) \right) \left( 4 + (3 \alpha-\delta) (\alpha+\delta) \right)} & \hspace{5mm} Q_e^2 = -V_0{ 2 - \delta (\alpha+\delta) \over 2 \left( 2 + \alpha (\alpha+\delta) \right)}  
\end{align}

The following three  conditions must be satisfied for this solution to be valid :
 \be\label{condf1}Q_e^2 >0 \Rightarrow \ { 2 - \delta (\alpha+\delta) \over 2 + \alpha (\alpha+\delta)}>0 \ee 
\be\label{condf2}  C_a^2>0 \ \Rightarrow   \left(2 + \alpha (\alpha+\delta) \right) 
\left( 4 + (3 \alpha-\delta) (\alpha+\delta)  \right)>0 \ee
\be\label{condf3} \gamma>0 \ \Rightarrow 1 -{ 2 \delta (\alpha+\delta) \over 4 + (\alpha+\delta)^2}>0 \ee
The last condition arises from the requirement that $g_{tt}$ vanish at the horizon, which we have taken to lie 
at $r=0$.

Now note that the constraint,   $Q_e^2>0$,  can be satisfied in general in two ways: both the numerator and the denominator 
are positive or both are negative. The latter possibility, however, it is easy to see, violates the condition $\gamma>0$. 
Therefore the conditions  above can be reexpressed as, 
\begin{eqnarray}\label{constraint1}
2 - \delta (\alpha+\delta)&>& 0 \\
\label{constraint2}
2 + \alpha (\alpha+\delta) &>& 0 \\
\label{constraint3} 4 + (3 \alpha-\delta) (\alpha+\delta) &>& 0.
\end{eqnarray}

Note that from eq.(\ref{case1}), (\ref{case11}) and eq. (\ref{constraint1}),
\be
\label{condgb}
\gamma-\beta={4-2\delta(\alpha+\delta)\over4+(\alpha+\delta)^2} >0.
\ee

The parameter $\beta+\gamma$ will play an important role in the subsequent discussion. From eq.(\ref{case1}), (\ref{case11}) it takes
the value,
\be
\label{bpg}
\beta+\gamma=1+{(\alpha+\delta)(\alpha-\delta)\over4+(\alpha+\delta)^2}.
\ee
Note that $\beta+\gamma=1$ when $\alpha=\pm \delta$. 
The case $\alpha+\delta=0$ has $\gamma=1,\beta=0$ and therefore corresponds to an
 $AdS_2\times R^2$ geometry which is also the near horizon geometry in the extreme RN case.
The case $\alpha=\delta$ has  $\beta>0$ and therefore corresponds to an extremal brane with vanishing horizon area. 

\subsection{More on the Solutions}
Here we comment on some properties of the solutions in more detail. 

The solution in eq.(\ref{case1}), (\ref{case11}) has only one parameter $V_0$, in particular the 
 charge too gets  fixed in it in  terms of this parameter.
In the full solution, including the asymptotic region near the boundary, the charge or the chemical potential would of course be an additional parameter,   however this parameter does not appear in the near-horizon solution. 
The solution above eq.(\ref{case1}), (\ref{case11}) is actually an exact  solution to the equations of motion, but  in a situation
where the asymptotic boundary conditions are different, say $AdS_4$, it will only be approximately valid at small values of $r$.
And  the chemical potential  will enter in the determination for when the near-horizon geometry 
stops being a good approximation
 \footnote{More generally, there could be additional scales, e.g., if a relevant operator is turned 
on in the boundary CFT, besides the chemical potential, to obtain the full geometry. In such a situation
our comments  apply if these additional scales are also of order the chemical potential. In Appendix F
we will in fact construct examples of such solutions where the relevant operator is dual to the dilaton.
For typical values of parameters considered there, the additional scale which corresponds to the coupling constant of this operator in the Lagrangian is of order $\mu$.}. 
 
If $V_0$ and $\mu$ are the only two scales in the geometry one expects that the near horizon geometry is a good approximation for 
\be
\label{apprr}
r\ll {\mu\over \sqrt{|V_0|}}.
\ee
Note that $r$ is dimensionless, $\mu$ and $\sqrt{|V_0|}$ have units of Mass \footnote{There is a hidden overall Newton constant $G_N$ in the action (\ref{genact}).}, thus this formula is consistent with dimensional analysis \footnote{Another way to obtain (\ref{apprr}) in the full solution viewpoint is to note that we need $r - r_h \ll r_h \sim \mu$ where $r_h$ is horizon. By restoring the length scale of the system $\sim 1/\sqrt{|V_0|}$ and by coordinate transformation so that we set horizon to be $r=0$, we obtain (\ref{apprr}).}.

Let us make a few more comments. 
When $\alpha+\delta$ does not vanish,  $\beta>0$,
and therefore the area of the horizon and thus the entropy vanishes. 
Second, the solution has a smooth limit when $\delta \rightarrow 0$, and reduces to the black brane found in 
\cite{GKPT},  \cite{MT} in this limit. Third,  the solution is somewhat analogous to Lifhsitz type solutions \cite{Lifchitzref},
however,  in general the  metric in the solution
 does not have any scaling symmetry. Exceptions arise    when $\gamma=1$, which requires
either $\alpha+\delta=0$, the eRN case mentioned above, or $\delta=0$, the  case studied in \cite{GKPT}. 
Finally, after a suitable coordinate transformation it is easy to see that the solution we have obtained above,
eq.(\ref{case1}), (\ref{case11}) agrees  with the solution discussed
in \cite{Kiritsis} in \S8, eq.(8.1a) - (8.1d), with the non-extremality parameter $m$ set to zero \footnote{For comparison purposes  the parameters $(\alpha,\delta)$ defined in this paper should be related to $(\gamma,\delta)$ in \cite{Kiritsis} as
follows: $(\alpha, \delta)\rightarrow (\gamma, -\delta)$.}.

We will examine the thermodynamics of the near extremal solution next and also comment on electrical conductivity. 

Before proceeding let us note that there is another class of solution to the equation of motion, also consistent 
with the ansatz, eq.(\ref{ansatz2}), which is valid in a region of the $(\alpha, \delta)$ parameter space which is
 different from the one described above. 
This class of solutions is discussed in Appendix A. A further study of this additional class of solutions is left for the 
future, the discussion   which follows will
mainly focus on  the extremal black brane solutions  discussed above.


\subsection{Thermodynamics of the Slightly Non-Extremal Black Brane}
Next we turn to constructing slightly non-extremal black brane solutions (these would have temperature $T\ll |\mu|$,
where $\mu$ is the chemical potential). 
The following observation makes it easy to do so. 
Starting from the extremal solution, of the form,  eq.(\ref{ansatz2}), 
the equations of motion continue to hold  if a term linear in $r$ is added to $a^2b^2$, while keeping $b^2$ and $\phi$
unchanged.  This is straightforward to see for the first three equations of motion, eq.(\ref{em1}-\ref{em3}),
and follows for eq.(\ref{em4}) after we note that it can be rewritten as,
\be
\label{em5}
\left(a^2b^2 {b'\over b}\right)'=-V_{eff}. 
\ee

This observation allows us to construct a one parameter deformation of the extremal solutions,
where 
\be
\label{onepa}
a^2=C_a^2r^{2\gamma}\left(1-\left({r_h\over r}\right)^{2\beta+2\gamma-1}\right)
\ee
and $b^2,\phi$ take the form in eq.(\ref{ansatz2}),
with $C_a, \gamma,\beta, k$ as given in eq.(\ref{case1}), eq.(\ref{case11}).  The parameter
$r_h$  characterizes the deformation and  corresponds to the location of the horizon. 
It is easy to see that the deformed solutions have a first order zero at the horizon and thus are non-extremal.
For $r_h\ll 1$ these solutions are close to extremal.

Near the horizon, 
\be
\label{nha}
a^2\simeq C_a^2 (2\beta+2\gamma-1)r_h^{2\gamma-1}(r-r_h)
\ee
We note that for the solution in eq.(\ref{case1}), (\ref{case11}), the condition, 
\be
\label{condt}
2\beta+2\gamma-1>0
\ee
is indeed met, as would be needed for the region $r>r_h$ to correspond to the region outside of the horizon.
It is  simple to see that the temperature of the non-extremal black brane goes like, 
\be
\label{Tg}
T\sim r_h^{2\gamma-1}
\ee
and the entropy density scales like, 
\be
\label{entropne}
s\sim r_h^{2\beta}\sim T^{2\beta\over 2\gamma-1}
\ee

A physically acceptable extremal black brane, which corresponds to the ground state of a conventional
 field theory on the boundary,
should have a positive specific heat when it is heated up. This leads to the additional condition 
for an acceptable solution,
\begin{equation}\label{thermoconstraint}
 2 \gamma -1 >0
\end{equation}
When expressed in terms of $\alpha,\delta$ this becomes, 
\begin{equation}\label{condspheat}
 4 + (\alpha - 3 \delta) (\alpha+ \delta) >0
\end{equation}
This condition must be added to the three discused earlier, eq.(\ref{constraint1}) - (\ref{constraint3}). 
In Figure 1  we show the region in the $(\alpha, \delta)$ plane which meets all these four conditions. 

\begin{figure}
\begin{center}
\includegraphics[scale=0.6]{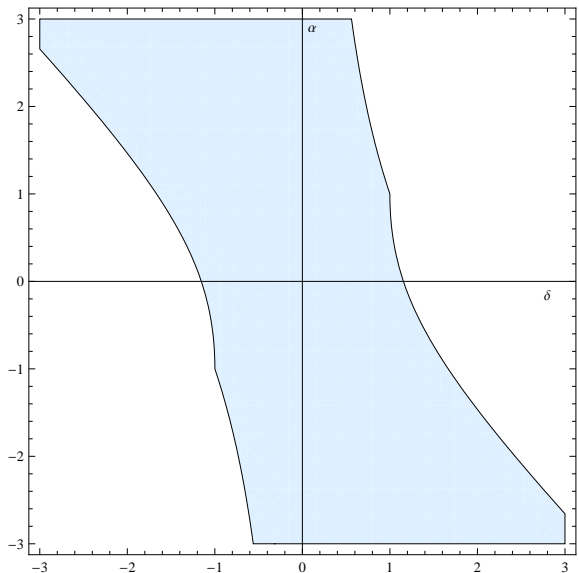}
\caption{Region Allowed by the Constraints}
\end{center}
\end{figure}

\subsection{Taming the Singularity}
In the discussion above we used classical  Einstein gravity and worked in the two-derivative approximation.
These approximations can break down  at sufficiently small values of the radial coordinate $r$.
For example, a curvature singularity could arise or the dilaton can diverge signalling such a breakdown.
 The essential point we want to make  in this subsection  is that turning on a  temperature which is very small
in the large $N$ limit  can often help control this breakdown. 
Our arguments are only suggestive at the moment,  a definitive discussion would require an embedding of these dilaton systems in string 
theory which has not been done as yet.

The parameter 
\be
\label{defL}
L={1\over\sqrt{|V_0|}}
\ee
is an important length  that characterizes the system. We will assume that $L$ and the chemical potential $\mu$ are the only two scales
in the system in the two derivative approximation.  For example, if the geometry is asymptotically $AdS_4$ the radius of
 $AdS_4$ would be of order $L$. 
From eq.(\ref{case1}), (\ref{case11}) we see that $L$ is also  the only scale in the near-horizon solution.
A measure of the number of degrees of freedom in the system is given by 
\be
\label{defN}
N^2={L^2\over l_{Pl}^2}.
\ee

In the near-horizon geometry of the extremal solution   the Ricci scalar, ${\cal R} \sim {r^{2(\gamma-1)} / L^2}$. 
The higher invariants $R_{\mu\nu}R^{\mu\nu}$ and $R_{\mu\nu\rho\sigma}R^{\mu\nu\rho\sigma}$ are also of the same order,
i.e. $R_{\mu\nu}R^{\mu\nu}\sim( {r^{2(\gamma-1)} / L^2})^2$ etc. We see that these invariants diverge at $r=0$ for $\gamma<1$ 
\footnote{For $\gamma\ge 1$ tidal forces could still blow up, as happens  in the Lifshitz solutions obtained when $\gamma=1, \beta\ne 0$.}. 
At finite temperature the divergence is cutoff at the horizon located at $r=r_h$. 
We get
\be
\label{curpu}
{\cal R} l_{pl}^2\sim {r_h^{2(\gamma-1)}\over N^2}
\ee
Expressing this in terms of the temperature  
\be
\label{tempn}
T\sim {r_h^{2\gamma-1} \over L} 
\ee
leads to 
\be
\label{nex}
{\cal R} l_{pl}^2\sim {(T L )^{2\gamma-2\over 2\gamma-1} \over N^2}
\ee
We see that for $N^2\gg 1$ the curvature can be made much smaller than the Planck scale  by taking 
\be
\label{condtemp}
\left({1\over N^2}\right)^{2\gamma-1\over 2(1-\gamma)} \ll T L \ll 1
\ee
Thus  the curvature can be made much smaller than the Planck scale, while keeping the temperature much smaller than $1/L$ . 
In the   large $N$ limit, 
where $N\rightarrow \infty$, keeping $L$ fixed, this condition is  in fact met for any non-zero temperature. 

However this analysis might be incomplete, since the 
$4$ dim. Planck scale is a derived quantity in string theory and the criterion for
 breakdown of classical two-derivative gravity involves the curvature in units of the string scale, which is  related to $l_{Pl}$ via the values of moduli,
 and also involves the string
coupling. It could be that requiring  the curvature to be much  smaller than the string scale imposes a stronger condition than 
eq.(\ref{condtemp}),
or that  a  stronger condition arises by requiring that quantum effects remain small \footnote{Another reason for thinking that 
there is more to this analysis is  the condition eq.(\ref{condtemp}) involves  $L$ which does not directly have an interpretation in the 
boundary theory.} .

For example, it could be that the dilaton  enters in the relation between the string and Planck scale \footnote{The scalar we are calling the dilaton may not literally be the dilaton field of string theory whose expectation value determines the string coupling.}, 
since the dilaton also varies with $r$
this could change the condition for the validity of the two-derivative approximation. 
Similarly, the dilaton might also enter in the string coupling and the 
 requirement that quantum corrections are small could impose  significant
restrictions. In fact this is likely to be the case.  
The gauge coupling in the action eq.(\ref{action}) goes like $g^2=e^{-2\alpha \phi}$.  One would  expect the theory to be weakly coupled only when
$g^2\ll 1$. The dilaton  in the near horizon region is given by, 
\be
\label{nhdil}
\phi=k\log(r/r_c)
\ee
where we have introduced a radial cut-off $r_c$ on the RHS. 
This leads to the condition, 
\be
\label{gc}
e^{-2\alpha \phi}=\left({r_h\over r_c}\right)^{-2\alpha k} \ll 1
\ee
When $\alpha k>0$ this does not allow  the temperature to become very small.  The parameter
$r_c$ depends on the chemical potential, which determines  how far out in $r$ the geometry begins to depart from the near-horizon solution,
and also can depend on the asymptotic value of $\phi$. 
It could be that this condition can be met only if $T$ is large compared to $\mu$,  this would require a temperature which is much too big,   the  resulting finite 
temperature black brane would not  be described by the near-horizon metric we have found.  Or it could be that if one starts
 with the asymptotic value of the dilaton being small enough this condition can be met while remaining within the scope of the solution
 we have found. 

Clearly, this question about the validity of our solution, in the presence of a small temperature,  
will need to be revisited in a more complete string construction.

\subsection{Conductivity}
For completeness let us also comment on the electrical conductivity of this system.
Our discussion follows \cite{GKPT} \S3 and \cite{GIKPTW} \S3 closely, we omit some details.
Before proceeding let us note that the compution of  the electrical conductivity in AdS/CFT is discussed in several other papers as well,
e.g., \cite{HKSS}, \cite{HR}, \cite{Leigh}, \cite{Jain}. 
The essential idea of our calculation, \cite{HR}, is to cast the equation governing  a perturbation of say the $A_x$ component
 of the gauge field in the black brane background in the form of a Schroedinger problem,
\be
\label{sp}
-{d^2\psi\over dz^2} +V(z) \psi = \omega^2 \psi
\ee
where $\omega$ is the frequency. 
Starting at the boundary with an ingoing pulse one can calculate the reflection amplitude $R$ from the potential $V(z)$.
The conductivity is then given  by 
\be
\label{condf}
\sigma={1-R\over 1+R}.
\ee
The dependence of the conductivity on $\omega, T$, for small values of these parameters
can be obtained, upto overall coefficients, by analyzing the behaviour of $V(z)$ in the near-horizon region. 
Thus our lack of knowledge of the full solution in the problem at hand will not be a 
limitation in extracting this information, although conceptually it is useful to assume that there is a screen eventually located in an asymptotically $AdS$ region.

Some more details are as follows. The variable $\psi$ is 
\be
\psi=f(\phi) A_x
\ee
and $z$ is given in terms of the radial variable  $r$  used in the metric, eq.(\ref{ansatz1}), by
\be
\label{defz}
\partial_z=a^2 \partial_r
\ee

As discussed in  appendix B, in the extremal near-horizon geometry, 
\be
\label{potext}
V(z)={c\over z^2}
\ee
where the coefficient $c$, which is  important for this calculation, takes the value, 
\be
\label{valc}
c=2{(4+\alpha^2-\delta^2)(4+(\alpha-2\delta)(\alpha+\delta))  \over (4+(\alpha-3\delta)(\alpha+\delta))^2}
\ee

Defining 
\be
\label{defnu}
\nu=\sqrt{c+{1\over 4}},
\ee
it then follows from the analysis of \S3 in \cite{GKPT} for example that the optical conductivity, for $\omega
 \ll\mu $  at zero temperature,
 is given by, 
\be
\label{opcond}
Re(\sigma)\sim \omega^{2\nu-1}\sim \omega^{2(4+\alpha^2-\delta^2)\over 4+(\alpha-3\delta)(\alpha+\delta)}
\ee
Similarly, the DC conductivity, for $\omega\rightarrow 0$, for small temperature, $T/\mu\ll 1$,  goes like,
\be
\label{dccond}
Re(\sigma) \sim T^{2\nu-1}\sim T^{2(4+\alpha^2-\delta^2)\over 4+(\alpha-3\delta)(\alpha+\delta)}
\ee
as follows from the analysis in \cite{GIKPTW} \S3 for example. 
There is in addition a delta function at zero frequency in $Re(\sigma)$ which we have omitted above. 

Note that using eq.(\ref{case1}), eq.(\ref{case11}), we can express the  exponent 
\be
\label{relnu}
2\nu-1={2 \gamma\over (2\gamma -1)}
\ee
Since $2\gamma-1>0$ from eq. (\ref{thermoconstraint}), the RHS is always positive
thus the exponent in both the  optical conductivity
and DC conductivity are positive. This means the optical conductivity  increases with increasing
frequency and the DC conductivity increases with increasing temperature, with the system behaving in effect as one with 
a ``soft gap'' \footnote{In a system with a conventional gap the conductivity would be exponentially sensitive to the temperature,
going like, $Re(\sigma) \sim e^{-\Delta\over T}$, instead of having the power-law behaviour we find.}.  As $(\alpha,\delta)$ are varied $2\gamma-1$ can become arbitrarily small (while remaining positive)
 and thus the exponent eq.(\ref{relnu}) can become very large so that the increase with frequency or temperature is very gradual.

The result for the optical conductivity agrees with \S8 of \cite{Kiritsis}. The DC conductivity does not agree. 
The answer above corresponds to the  DC conductivity as defined by 
the two-point current -current  correlation function
using the Kubo formula. The definition in \cite{Kiritsis}, \S5, for the DC conductivity is different and related to
 the drag force on a massive charge.

\section{Fermionic Two Point Function}
We will consider a free fermion in the bulk with mass $m$ and charge $q$. Its action is 
\be
\label{actff}
S_{fermion}=\int d^{3+1}x \sqrt{-g}i[\bar{\psi}\Gamma^MD_M\psi-m\bar{\psi}\psi]
\ee
We will mostly follow the spinor and related Dirac matrix notation of \cite{MIT} and specifically comment on any 
differences below.
In our notation then, 
\be
\label{nota}
\bar{\psi}=\psi^\dagger \Gamma^{\bar{t}},  D_M=\partial_M+{1\over 4} \omega_{abM} \Gamma^{ab}-iq A_M
\ee
where $\omega_{abM}$ is the spin connection and $A_M$ is the vector potential. 
The gamma matrices,
\be
\label{convgamma}
\Gamma^r=\begin{pmatrix} 1 & 0 \cr 0 & -1 \cr \end{pmatrix}, \Gamma^\mu=\begin{pmatrix} 0 &  \gamma^\mu \cr \gamma^\mu &  0 \cr 
\end{pmatrix}
\ee
with 
\be
\label{defgamma}
\gamma^0=i\sigma_2, \gamma^1=\sigma_1, \gamma^2=\sigma_3.
\ee
where $\sigma_i, i=1, \cdots 3$ denote the Pauli matrices.

The spinor $\psi$ has four components and we define $\psi_{\pm}$ to be the upper and lower two components respectively,
\be
\label{defpsi}
\psi =\left(\begin{array}{c} \psi_+\\\psi_- \end{array} \right)
\ee
Using the translational symmetries we can take, 
\be
\label{defphi}
\psi_{\pm}=(-gg^{rr})^{-{1\over 4}}e^{-i\omega t + i k_ix^i} \phi_{\pm}
\ee

Asymptotically,
\be
\label{asforma}
\phi_+=Ar^m+Br^{-m-1}, \phi_-=Cr^{m-1}+Dr^{-m}
\ee
where $A,B,C,D $ are 2 dim. column vectors. 
If we define $D=SA$ then The two-point function on the boundary for the dual operator is 
\be
\label{bndrygf}
G_R=-iS\gamma^0
\ee

We can simplify the calculations by using the rotational symmetry in $x,y$ plane to choose $k=k_1$. 
Then writing
\be
\label{defyz}
 \phi_{\pm}= \begin{pmatrix} y_\pm \cr z_\pm \cr \end{pmatrix}
\ee
The equation of motion for $\psi$ then breaks up into block $2\times 2$ form coupling only 
$(y_+,z_-)$ and $(y_-,z_+)$ together respectively.   
The two-point function $G_R$ has two non-vanishing components, $G_{R11}$ and $G_{R22}$, these are related by, 
$G_{R11}(\omega,k_1)=G_{R22}(\omega,-k_1)$ and are not independent.  In what follows we will set   
 $(y_-,z_+)$ to  vanish, this is sufficient to extract $G_{R22}$ and then also $G_{R11}$. 

Also it will be convenient to make one change of variables and work with
\be
\label{defzp}
z_-'=i z_-
\ee
instead of $z_{-}$. 

  The equations for $(y_+,z_{-}')$ take the form, 
\begin{eqnarray}
\label{yfirsto} \sqrt{g_{ii}\over g_{rr}}\left(\partial_r - m\sqrt{g_{rr}}\right)y_{+} &=& -(k_{1}-u)z_{-}'\\
\label{zfirsto}\sqrt{g_{ii}\over g_{rr}}(\partial_r + m\sqrt{g_{rr}})z_{-}' &=& -(k_{1}+u)y_{+}
\end{eqnarray}
with
\be
\label{defu}
u=\sqrt{g_{ii}\over -g_{tt}}(\omega+qA_t)
\ee

Asymptotically, towards the boundary \footnote{We are assuming here that asymptotically the spacetime approaches $AdS_4$, as happens
for the examples considered in appendix F. 
While this is useful for extracting the field theory results,  at low-frequency the essential features in these
 results  really only depend 
 on the near-horizon region.}
,  it is easy to see that the solution take the form, 
\be
\label{formas}
\left(\begin{array}{c} y_+\\ z_-'\end{array}\right) = C_1 \left(\begin{array}{c}  1 \\ 
 -{(\omega+\mu q + k_1)\over (2m-1) r} \end{array} \right) r^m + C_2 \left(\begin{array}{c} {-(\omega+\mu q-k_1)\over (2m+1) r}\\1\end{array} \right)
r^{-m}
\ee
where $\mu $ is the asymptotic value of the gauge potential $A_t$. 
Comparing with eq.(\ref{asforma}) and eq.(\ref{bndrygf}) we find that 
\be
\label{gr22} G_{R22}=-{C_2\over C_1}.
\ee

In the near horizon region, eq.(\ref{case1}), (\ref{case11}) the fermion equations of motion, eq.(\ref{yfirsto}), (\ref{zfirsto}) take the
form,
\begin{eqnarray}
\label{ya} r^{\beta+\gamma} \left(\partial_r - {m\over C_a r^\gamma}\right)y_{+} &=& - \left({k_{1}\over C_a} 
-{r^{\beta-\gamma}\over C_a^2}(\omega+qA_t) \right)z_{-}'\\
\label{za} r^{\beta+\gamma} \left(\partial_r + {m\over C_a r^\gamma}\right)z_-' &=& - \left({k_{1}\over C_a} 
+{r^{\beta-\gamma}\over C_a^2}(\omega+qA_t) \right)y_+
\end{eqnarray}
The explicit $C_a$ dependence will vanish if we work with the rescaled variables,  $m= \tilde{m} C_a, k_1= \tilde{k_1} C_a, 
(\omega+qA_t)\rightarrow C_a^2(\tilde{\omega}+\tilde{q} A_t)$. To avoid clutter we will refer to these rescaled variables
$\tilde{m},\tilde{k_1},\tilde{\omega},\tilde{q}$ as $m,k_1,\omega,q$ in the discussion below. 
Eq.(\ref{ya}) and (\ref{za}) now take the form, 
\begin{eqnarray}
\label{yfirst} r^{\beta+\gamma} \left(\partial_r - {m\over  r^\gamma} \right)y_{+} &=& - \left(k_{1} 
-r^{\beta-\gamma}(\omega+qA_t) \right)z_{-}'\\
\label{zfirst} r^{\beta+\gamma} \left(\partial_r + {m\over r^\gamma} \right)z_-' &=& - \left(k_{1} 
+r^{\beta-\gamma}(\omega+qA_t) \right)y_+
\end{eqnarray}

We will be interested in the retarded Green's function in the bulk, 
this is obtained by imposing in-going boundary conditions at the horizon.
Very close to the  horizon where $\omega$ term dominates, eq.(\ref{yfirst}), (\ref{zfirst}) become, 
\begin{equation}
\label{nheq}
r^{\beta+\gamma} \ \partial_r \left( \begin{array}{c} y_{+} \\ z_{-}^{'}  \end{array} \right) = \omega r^{\beta-\gamma} \ i \sigma_{2}\left( \begin{array}{c} y_{+} \\ z_{-}^{'}  \end{array}\right)
\end{equation}
The ingoing solution is obtained by taking
\begin{equation}
\label{nhpsi}
\left( \begin{array}{c} y_+ \\z_{-}' \end{array} \right) =
 \left(\begin{array}{c} 1 \\- i \end{array} \right) e^{-i\omega z}
\end{equation}
where 
\be
\label{defzvch}
z={ 1 \over (1-2\gamma)r^{2\gamma-1}}
\ee
Note that the time dependence has been taken to be of form $e^{-i\omega t}$, eq.(\ref{defphi}),
 and since 
$z \rightarrow -\infty$, at the horizon,  where $r\rightarrow 0$,
$e^{-i\omega(t+z)}$ is well behaved at the future horizon where $t\rightarrow \infty$.

We are interested in the  small frequency behaviour of the boundary two-point function.
At a Fermi surface, where $k_1=k_F$,  the boundary two-point function has a singularity,   for $\omega \rightarrow 0$.
We will be interested in asking whether such a surface can arise in this system and what is the nature of small frequency excitations near this surface. 
It will be convenient to divide our analysis into three parts, depending on the value the parameter $\beta+\gamma$ takes.

When $\beta+\gamma>1$ we will see that the boundary fermionic two-point function is of Fermi liquid type. 
More correctly, as was discussed in the introduction the small frequency excitations have a linear dispersion relation, with a width which is narrower than $\omega^2$.
When $\beta+\gamma<1$ we will find that the low-frequency excitations acquire a width which is  
  non-vanishing  even in  the $\omega\rightarrow 0$ limit, and thus is very broad.  
The transition region,  $\beta+\gamma=1$ consists of two lines.
One of them corresponds to extremal RN type geometries, which are well known to give Non-Fermi liquid behaviour  \cite{SSL}, \cite{MIT}, 
\cite{CSZ}. The other corresponds to geometries which have vanishing entropy, here we find that the behaviour can be of both Fermi or non-Fermi liquid type with width $\Gamma \sim \omega^p$. The power  $p>0$ and  can be bigger, equal to, or less than two, so that 
one can get both Fermi-liquid and non-Fermi liquid behaviour. 

We now turn to discussing these three cases in turn. 
In Figure 2 we plot the regions where $\beta,\gamma$ take different values, in the $(\alpha,\delta)$ plane. 

\begin{figure}
\begin{center}
\includegraphics[scale=0.6]{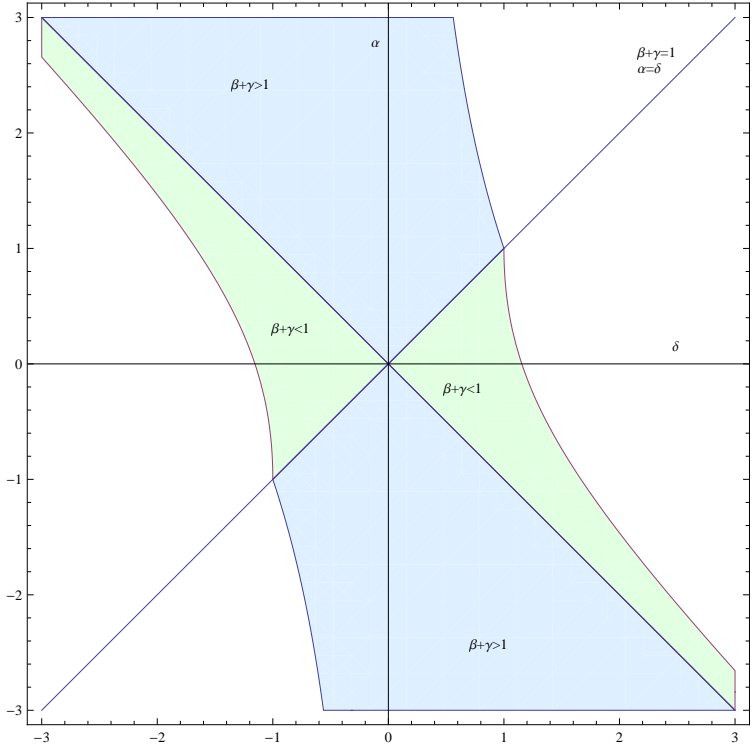}
\caption{Region with $\beta+\gamma>1$  in Blue;
$\beta+\gamma<1$ in Green.}
\end{center}
\end{figure}

Before proceeding let us comment on the significance of the parameter $\beta+\gamma$.
We  denote \footnote{In the action  eq.(\ref{action}) etc above  $\psi$ stands for the four components fermion, in the subsequent 
discussion we will only be exciting
the $(y_+,z_-')$ components and to save clutter henceforth  will refer to the two component spinor itself  as $\psi$. }
\be
\label{defpsinew}
\psi= \left(\begin{array}{c} y_+\\z_-' \end{array}\right).
\ee
There is a convenient way to recast eq.(\ref{yfirst}) and (\ref{zfirst}) as a second order equation of Schrodinger form, with a spin dependent potential:
\begin{eqnarray}
\label{secondorrr}
r^{\gamma+\beta}\partial_r(r^{\gamma+\beta}\partial_r\psi) & = & [m^2r^{2\beta}+k_1^2-(\omega+qA_t)^2r^{2(\beta-\gamma)}] \psi
 +\beta m r^{2\beta+\gamma-1}\sigma_3 \psi \\ \nonumber &&
 +  i [(\beta-\gamma)r^{2\beta-1}(\omega+q A_t)+q r^{2\beta} \partial_rA_t]\sigma_2 \psi  \,.
\end{eqnarray}
The Shrodinger variable  is 
\be
\label{defzeta}
\zeta=\int {dr \over r^{\gamma+\beta}} = {r^{1-\beta-\gamma} \over (1-\beta-\gamma)}.
\ee
The distance as measured in this variable to the horizon ($r=0$) has a power-law divergence for $\beta+\gamma>1$, a logarithmic
divergence for $\beta+\gamma=1$ and no divergence for $\beta+\gamma<1$.
This difference yields the interaction strength difference between bulk fermion and black brane. 
Depending on the distance to the horizon being infinite or finite, it is expected that  
the influence of the black brane to the bulk fermion is highly suppressed or large. 
Since at the horizon the bulk fermions decay into black brane degrees of freedom,  
this distance difference ultimately results in the three qualitatively different types of behaviour (decay ratio width) for the boundary fermions two-point function as we will see below.

Eq.(\ref{secondorrr}) is convenient for our analysis for the following reason. 
We are interesting in the  dominant $\omega$ dependence at small  frequency. 
As we will see, this dependence arises   from a  region close to the horizon where  
the $m,A_t$ dependent terms on the RHS of eq.(\ref{secondorrr}) can be neglected.   This only leaves the $k_1$ and $\omega$ dependent terms 
in eq.(\ref{secondorrr}). 
Of these the $k_1$ dependent term is   particularly important, since it is repulsive and hinders the particle from falling into the 
horizon. Physically this is because the shrinking size of the $x,y$ directions  gives rise to  a
 cost in the $k_1$ dependent energy that increases closer to the horizon.  In eq.(\ref{secondorrr}) this is seen clearly since the 
$k_1$ dependent term does not give rise to a spin dependent potential (which would be proportional to the $\sigma$ matrices)
and only multiplies the identity matrix in spin space, with a sign which corresponds to its providing positive potential energy.


\subsection{$\beta+\gamma>1$}
In this case we will see that the WKB approximation can be used to calculate the boundary Green's function. 
The width of  the boundary correlator is related to  tunneling through a classically disallowed region in the bulk
and will be exponentially suppressed at small frequency.

In the WKB approximation radial derivatives on $(y_+,z_-')$ are more important than those on the metric or gauge potential,
From eq.(\ref{yfirst}) and eq.(\ref{zfirst}) it then follows that for example $y_+$ satisfies the equation \footnote{This equation 
has additional terms which are small in the WKB approximation.},
\begin{equation}
\label{wkba}
r^{\beta+\gamma}\partial_r (r^{\beta+\gamma}\partial_r y_+)  -m^2 r^{2\beta} y_+ - \left( k_1^2-r^{2(\beta-\gamma)}
(\omega+q A_t)^2\right)y_+ =0. 
\end{equation}

We will be mainly interested in the small frequency behaviour, which is essentially determined by the near horizon region, where
\be
\label{condr2}
r\ll 1.
\ee
In this region we see that the mass dependent terms are subdominant compared to the $\omega$ and $k_1$ dependent terms. 
Also, from eq.(\ref{gf}) we see that in this region the gauge potential is given by, 
\be
\label{gpot}
A_t\sim r^{1-2\alpha k-2\beta}.
\ee
From eq.(\ref{case1}), (\ref{case11}) we see that for $\beta+\gamma>1, (\alpha + \delta)(\alpha - \delta)>0$. It will turn out that 
 the small frequency behavior
can in fact be extracted from the region where 
\be
\label{condr1}
r^{1-(2\beta+2\alpha k)}\ll \omega
\ee
so $A_t$ can also be neglected compared to the $\omega$ dependence \footnote{We take $q\sim O(1)$.} in eq.(\ref{wkba}).  
Note that $1- (2\alpha k+2\beta)  > 0$ under eq. (\ref{condspheat}), therefore eq. (\ref{condr2}) and (\ref{condr1}) are compatible. 

The equation eq.(\ref{wkba}) can   be cast in the form of a Schroedinger equation for a zero energy eigenstate,
\be
\label{sea}
-{d^2 y_{+} \over {d \zeta^2}} +  V y_+=0
\ee
where  $\zeta$ is  defined in eq.(\ref{defzeta}),
and $V$, the potential, is 
\be
\label{defpot}
V= k_1^2 -{\omega^2\over r^{2(\gamma-\beta)}} 
\ee

There are three regions of interest in the calculation. The region very close to the horizon, which we will call $R_1$
is where $r\rightarrow 0$.
This region is classically allowed, as  follows from  eq(\ref{condgb}) which imply that   
 $\gamma-\beta>0$, therefore $V < 0$. 

The second region which we call $R_2$ is where 
\be
\label{sr}
1\gg r^{{\gamma-\beta}}\gg { \omega \over k_1}
\ee
so that the $k_1$ term dominates over the $\omega$ dependent term in $V$.
Note we  still have to meet eq.(\ref{condr1}); at the end of this section we will return to this point and show that eq.(\ref{condr1})
and eq.(\ref{sr}) are indeed compatible. 
Note that the $R_2$ region is classically disallowed and in this region the frequency dependence is unimportant. 

Finally the third region $R_3$ is close to the boundary where $r\rightarrow \infty$. 

The  regions $R_1$ and $R_2$  are separated by a turning point at 
\be
\label{tp}
r_{tp}= \left({\omega \over k_1}\right)^{1\over {\gamma-\beta} }
\ee
It is easy to see that in $R_1$
 the solution to eq.(\ref{sea}) in the WKB approximation, which corresponds to in-going boundary conditions at the horizon, is 
\be
\label{ingwkb}
y_+={1\over \sqrt{\hat{k}}} e^{-i\int  \hat{k}(r) dr}
\ee
where
\be
\label{defhatk}
\hat{k}(r)={1 \over r^{2\gamma}}\sqrt{\omega^2-k_1^2 r^{2(\gamma-\beta)} }.
\ee
and the constant in the integration is chosen so that 
$\int \hat{k} dr \simeq {\omega \over (1-2\gamma) r^{2\gamma-1}}$ for $r\rightarrow 0$. 

In region $R_2$  there are two independent solutions to eq.(\ref{sea}) which  in the WKB approximation go like 
\be
\label{fpm}
f_{\pm}= e^{\mp \left( {k_1\over (\gamma+\beta-1)r^{(\gamma+\beta-1)}} \right) }.
\ee

Matching to eq.(\ref{ingwkb}) using standard turning point formulae, see, e.g, \cite{Migdal}, gives, 
\be
\label{fyp}
y_+={A\over \sqrt{\hat{k}}}[f_+ + {i\over 2} e^{-2I} f_-]
\ee
where  now
\be
\label{newkhat}
\hat{k}={k_1\over r^{\gamma-\beta}} \,.
\ee
and,
\be
\label{valI}
I=c_1 \left( {k_1^{2\gamma-1}\over \omega^{\beta+\gamma-1}}\right)^{1\over\gamma-\beta} \,,
\ee
\be
\label{valc1}
c_1=\int_1^\infty {dx \over x^{2\gamma}}\sqrt{x^{2(\gamma-\beta)}-1} \,,
\ee
and $A$ is an overall coefficient. 
Notice that $e^{-2I}$ is exponentially suppressed when $\omega\rightarrow 0$. 
Thus the  $f_-$ term has a very small coefficient and is subdominant at small $\omega$. 

From eq.(\ref{yfirst}), (\ref{zfirst}) it is easy  to find the  solution for $z_-'$ in this region. 
The result can be stated as follows. There are two linearly independent solutions in  region $R_2$,  
\be
\psi_{\pm}= \left(\begin{array}{c} 1 \\ \mp 1 \end{array}\right)f_{\pm}.
\ee
The solution  which agrees with eq.(\ref{fyp}) is 
\be
\label{fullsol1}
\psi={A\over \sqrt{\hat{k}}}[\psi_++{i\over 2} e^{-2I} \psi_-]
\ee 
Note that in this solution the  frequency dependence is summarized in the coefficient $e^{-2I}$, the solutions, $\psi_{\pm}$ are independent of
frequency.   

Now $\psi_{\pm}$  can further be extended from region $R_2$ to region $R_3$ which lies close to the boundary. 
Let the coefficients $C_1,C_2$, eq.(\ref{formas}),  which arise from $\psi_{\pm}$ be denoted by $C_{1\pm}, C_{2\pm}$  respectively. 
Then the  boundary two-point function is given by 
\be
\label{btp}
G_{R22}=-{C_{2+} + {i \over 2} e^{-2I} C_{2-} \over C_{1+}+{i\over 2 } e^{-2I}C_{1-}}
\ee
A Fermi surface arises when the coefficient $C_{1+}$ vanishes for $\omega\rightarrow 0$. 
In general this imposes one  real condition on the momentum $k_1$ and for a suitably chosen $k_1=k_F$ and if necessary
 by adjusting the geometry \footnote{One can also vary the dilaton dependence of the gauge coupling and potential once the dilaton is not in the
run-away region.} for 
the spacetime which interpolates between the near horizon region  and the $AdS$ boundary etc
 it should be possible to meet 
this condition.  By rotational invariance this will then be true for all $|\vec{k}|=k_F$. 
Expanding $C_1$ near $k=k_F$ then gives, 
\be
\label{ttp}
G_{R22}={c_3\over \omega - v_F (|\vec{k}|-k_F) + i c_2 e^{-2I}}
\ee
$v_F$ and $c_2$ arise from the Taylor series expansion of $C_{1+}$ and the leading behaviour of $C_{1-}$, 
we have neglected the term proportional to $C_{2-}$ in the numerator, and the leading $C_{2+}$ dependence feeds into the numerator 
$c_3$. 

We see that the small frequency excitations have a linear dispersion relation, with a width given by, 
\be
\label{width}
\Gamma\sim e^{-2I}=exp\left[-2c_1\left({k_F^{2\gamma-1}\over \omega^{(\beta+\gamma-1)} }\right)^{1\over \gamma-\beta}\right].
\ee
This width  is exponentially suppressed at small frequency and therefore very narrow.

Let us end by checking the validity of our approximations.  
Our use of the WKB approximation in regions $R_1, R_2$   imposes restrictions. 
This approximation requires that radial derivatives acting on $\psi=\left(\begin{array}{c}y_+ \\ z_-' \end{array}\right)$ 
are more important than derivatives of the metric. 
In region $R_2$ this gives the condition
\be
\label{condf222}
r^{\gamma+\beta-1}\ll 1
\ee
(we have set $k_1\sim O(1)$).
Note that  eq.(\ref{condf222}) can be met in the near horizon geometry where $r\ll 1$ only if
$\beta+\gamma>1$. The fact that the WKB  approximation breaks when $\beta+\gamma\le 1$   is also suggested by the decay width
eq.(\ref{width}) which is no longer suppressed at small $\omega$.
In  addition, we have assumed that eq.(\ref{condr1}) is correct so that the gauge potential dependent  terms can be dropped.
In region $R_2$ this has to be compatible  with the condition eq.(\ref{sr}). One can show from eq.(\ref{case1}), eq.(\ref{case11}) 
that in the region where $\beta+\gamma>1$, $\gamma-\beta<1-(2\beta+2\alpha k)$, it therefore follows that for small $\omega$
these two conditions are compatible. 

In Region $R_1$ far from the turning point validity of WKB approximation requires, 
\be
\label{condfa}
r\ll \omega^{{1\over 2\gamma-1}}
\ee
In addition eq. (\ref{condr1}) needs to be met. 
These are clearly compatible, in fact eq.(\ref{condfa}) is more restrictive.

\subsection{Some General Comments for the Cases $\beta+\gamma\le 1$}
We will now turn to analyzing what happens when $\beta+\gamma\le 1$. 

A few general comments are worth making before we go into details. 
From eq.(\ref{case1}), eq.(\ref{case11}) we see that $\beta+\gamma=1$ corresponds to the lines, 
$\alpha=-\delta$ and $\alpha=+\delta$. The first case, $\alpha=-\delta$ corresponds to an $AdS_2\times R^2$ metric 
which is   the near-horizon geometry of the extreme
Reissner Nordstrom Black Brane.  This has been analyzed extensively in \cite{SSL} - 
\cite{MIT4}, and we will not elaborate on this case further. 
The second case, $\alpha=\delta$, necessarily has $\beta\ne 0$ (for $\alpha\ne 0$ ), it is not $AdS_2\times R^2$ 
and  has vanishing area. 

In the extremal RN case while studying the fermion equation of motion  eq.(\ref{yfirst}), eq.(\ref{zfirst}) at small frequency
  the dependence on $m,k_1$ and charge through $A_t$ dependent terms  are all important.  
In contrast for the $\alpha=\delta$ case and for all cases where $\beta+\gamma<1$ both the 
$A_t$ and $m$ dependent terms in eq.(\ref{yfirst}), (\ref{zfirst}) can be neglected, in the near-horizon region
relevant for determining the small frequency behaviour. This results in considerable simplification of the analysis. 

It turns out that  extracting the $\omega$ dependence  requires us to solve the equations from the horizon 
upto  a radial location where
\be
\label{limra}
1\gg r \gg \omega^{1\over 2\gamma-1}.
\ee
Beyond that the $\omega$ dependence  turns out to be  subdominant and can be neglected \footnote{
One expects on general grounds that the gravitational redshift 
is monotonic as one goes from the black brane horizon to the boundary  making the $\omega$ dependence increasingly negligible.} .
Now from eq.(\ref{yfirst} \ref{zfirst}) we see that  
the  $A_t$ dependence  is unimportant  if $|A_t|\ll \omega$. From eq.(\ref{condr1}) this leads to the condition, 
\be
\label{inte}
r\ll \omega^{1\over 1-2\alpha k-2\beta}
\ee
which is compatible with eq.(\ref{limra}) if  
\be
\label{condatb}
\omega^{1\over 2\gamma-1}\ll \omega^{1\over 1-2\alpha k-2\beta} .
\ee
This last condition  is true for $\omega\ll 1$  because eq.(\ref{case1}), eq.(\ref{case11}) imply that $2\gamma-1 < 1-2\alpha k-2\beta$ for 
$\delta\ne -\alpha$.

The $m$ dependent terms  can be neglected if they are small  compared to the effect of $\partial_r$.
Now $\psi$ will be vary at least  as rapidly as a power of $r$, 
from eq.(\ref{yfirst}), (\ref{zfirst}) the condition for neglecting the $m$ dependent term then  becomes 
\be
\label{mdepcond}
{m\over r^\gamma}\ll {1\over r}.
\ee
For $m\sim O(1)$, $r\ll 1$  this gives, $\gamma<1$, which is true since $\beta + \gamma \le 1$ and $\beta>0$ when $\alpha\ne -\delta$. 

Henceforth we will study the cases $\beta+\gamma<1$ and the branch $\alpha=\delta$ for the case $\beta+\gamma=1$
and  therefore  we can set the $A_t,m$ dependent terms to be zero in \footnote{To ensure clarity
let us reiterate that below when we refer to $\beta+\gamma=1$ we only mean the case where $\alpha=\delta$,
for which  the metric is not $AdS_2\times R^2$.} eq.(\ref{yfirst}), eq.(\ref{zfirst}). 

Eq.(\ref{yfirst}), (\ref{zfirst}) then become, in terms of 
\be
\label{defpsin}
\psi=\left(\begin{array}{c} y_+\\z_-'\end{array}\right),
\ee
\be
\label{eomn}
r^{\beta+\gamma}\partial_r\psi=(-k_1\sigma_1+i r^{\beta-\gamma} \omega \sigma_2) \psi
\ee

The behaviour of the solution can be understood  qualitatively as follows. 
Very close to the horizon, the $\omega$ dependent term on the RHS will dominate over the $k_1$ dependent one since 
$\beta - \gamma < 0$ as eq. (\ref{condgb}).
Thus $\psi$ will be of the form given in eq.(\ref{nhpsi}). The effects of the frequency will become
subdominant to $k_1$ dependent ones when the $\omega$ dependent term on the RHS of eq.(\ref{eomn}) becomes less important
compared to the $k_1$ dependent term giving the condition, 
\be
\label{condk1}
r\gg \left({\omega\over k_1}\right)^{1\over \gamma-\beta}.
\ee

Now  another way to estimate when the effects of frequency become small is when the 
the phase in eq.(\ref{nhpsi}) becomes small. Using eq.(\ref{defzvch}) this gives
\be
\label{psm}
|\omega z|\sim |{\omega \over r^{2\gamma-1}}|\ll 1
\ee
which implies, 
\be
\label{rcondc}
r\gg \omega^{1\over 2\gamma-1}
\ee

Now it is easy to see that for $\beta+\gamma<1$, $\omega^{1\over 2\gamma-1}<\omega^{1\over \gamma-\beta}$ for 
$\omega \ll 1$.
Thus as $r$ is increased from the horizon eq.(\ref{rcondc}) will be met before eq.(\ref{condk1}) is met \footnote{We are assuming
$k_1$ is $O(1)$ here.}.  For the case when $\beta+\gamma<1$ then in the region where
\be
\label{reginta}
\omega^{1\over 2\gamma-1}\ll r \ll \left({\omega\over k_1}\right)^{1\over \gamma-\beta}
\ee
the  solution can be obtained by simply expanding the exponential in eq.(\ref{nhpsi}) and gives, 
\be
\label{ppsi}
\psi =\left(\begin{array}{c} 1 \\-i \end{array}\right) + O(\omega z).
\ee
Going to large values of $r$  the $k_1,m$ and $A_t$ dependence will become important, but in this region the $\omega$ dependence
 can be neglected. 
We will return to studying the consequences of our  analysis above for the $\beta+\gamma<1$ case  in \S3.4. 

In contrast, for the $\beta+\gamma=1$   case the two exponents in eq.(\ref{condk1}) and (\ref{rcondc}) are the same, 
since $\beta=1-\gamma$. Thus, the $k_1$ dependence will become important before the phase factor in eq.(\ref{nhpsi}) can be 
approximated to unity.  Going to large values the $k_1$ dependent term will be more important and the solution will take the form, 
\be
\label{form2}
\psi=d_1 \left(\begin{array}{c} 1 \\-1 \end{array}\right) r^{k_1}+ d_2 \left(\begin{array}{c} 1 \\1 \end{array}\right) r^{-k_1}
\ee
At even large values of $r$ the $A_t,m$ dependent terms will also get important. 
We will  turn to a more complete analysis of this case in \S3.3. 

\subsection{More on the   $\beta+\gamma=1$ case}
Here we will be interested in solving the fermion equation in the background,  eq.(\ref{case1}), eq.(\ref{case11}),
with  $\alpha=\delta$.  
 
Our starting point is eq.(\ref{eomn}). 
It is  convenient to define,
\be
\label{defxipm}
\chi_{\pm}= y_+\pm z_-'.
\ee
And work with the variable, 
\be
\label{deftz}
\tilde{z}={\omega\over (2\gamma-1) r^{2\gamma-1}},
\ee
which goes to infinity at the horizon.
For now we specialize to the case when $k_1>0$ and define 
\be
\label{defeta}
\eta={k_1\over 2\gamma-1}.
\ee 
Then  eq.(\ref{eomn}) becomes, 
\begin{eqnarray}
\label{chieqn1}
 (\tz \partial_\tz - \eta) \chi_+ &=&  \tz \chi_- \\
\label{chieqn2}
  (\tz \partial_\tz + \eta) \chi_- &=& - \tz \chi_+,
\end{eqnarray}
which gives the second order equation, 
\begin{eqnarray}
 \left[ \tz^2 \partial_\tz^2 + (\tz^2 -\eta^2 -\eta) \right] \chi_- &=& 0.
\end{eqnarray}

The solution  for $\chi_-$ with the ingoing boundary conditions at the horizon is  then
\be
\label{solxim}
\chi_- = \sqrt{\tz} H^{(1)}_{{1 \over 2} +\eta } (\tz),
\ee
and from eq.(\ref{chieqn2}) for $\chi_+$ is 
\begin{equation}
\label{solxip}
\chi_+ = - \tz^{-\eta} \partial_\tz (\tz^{\eta} \chi_-).
\end{equation}

As discussed in Appendix D in the region where $\tz \ll 1$, i.e., $\omega\ll r^{2\gamma-1}$, the solution  
for $\psi=\left(\begin{array}{c} y_+ \\ z_-' \end{array}\right)$, in terms of the radial variable $r$ 
and upto an overall $\omega$ dependent normalization which is not important, becomes   
\be
\label{finalsola}
\psi=\left(\begin{array}{c}1\\-1\end{array}\right) r^{k_1} +  d e^{i\phi} \omega^{2\eta} \left(\begin{array}{c}1\\1\end{array}\right)r^{-k_1}.
\ee
Here $d$ and the phase $e^{i\phi}$,  which  in general has both a real and imaginary part,
 depend on $\gamma,\beta,k_1$,   but are independent of $\omega$.

Actually, as noted in Appendix D, this solution is only valid when $\eta<1/2$. For the case $\eta>1/2$ the solution takes the form
\be
\label{finalsola2}
\psi=\left(\begin{array}{c}1\\-1\end{array}\right) r^{k_1} +  i d'  \omega^{2\eta} \left(\begin{array}{c}1\\1\end{array}\right)r^{-k_1},
\ee
so that the second term  on the RHS makes a contribution only to the imaginary part of $\psi$. 
There are subleading corrections on the RHS  to the real part of $\psi$ which are suppressed by a power of $\omega$, these are not 
kept in eq.(\ref{finalsola2}) since their effect on the Green's function is  comparable to $\omega$ dependent contributions generated
when evolving the solution further out towards the boundary.

Note that the result eq.(\ref{finalsola}), eq.(\ref{finalsola2}) agrees with eq.(\ref{form2}) above.
It is also worth commenting
 that the result eq.(\ref{finalsola}), eq.(\ref{finalsola2}) agrees with what one would get by continuing the WKB results of \S3.1 to the case
$\beta+\gamma=1$. More precisely, from eq.(\ref{defzeta})  we see that for 
 $\beta+\gamma=1$, $\zeta =\log r$. As a result for $r\gg r_{tp}$
the exponential factor in the two solutions go like \footnote{
There is also a prefactor ${1\over \sqrt{\hat{k}}}$ which gives rise to an 
additional powerlaw in $r$. We are not keeping this term.} $r^{\pm k_1}$ which agrees with eq.(\ref{finalsola}), eq.(\ref{finalsola2}). The solution which grows, as $r$ increases, 
has wavefunction $\left(\begin{array}{c} 1\\-1\end{array}\right)$ in the two component 
$\left(\begin{array}{c} y_+ \\ z_-' \end{array}\right)$
space and the falling solution
has wave function $\left(\begin{array}{c} 1\\1\end{array}\right)$, just as in the WKB case. The only difference is that the 
WKB suppression factor which was exponentially suppressed in $\omega$ has now turned into a power-law suppression in eq.(\ref{finalsola}),
eq.(\ref{finalsola2}).
This is because the fermion wave function can now penetrate the barrier more easily and thus can have a bigger mixing
 with the modes in the vicinity of the horizon. This crossover from the exponential suppression to a 
power-law has also been discussed in  \cite{FP}.

So far for ease of discussion we have considered the case when $k_1>0$. For the case $k_1<0$ a very similar
 analysis can be carried out, as discussed in Appendix D. Let us define $\eta$, both for the case when $k_1>0$ and  $k_1<0$ to be
\be
\label{valeta}
\eta={|k_1|\over 2\gamma-1},
\ee
so that $\eta>0$. 
Now for $k_1<0$ and $\eta<1/2$, the solution eq.(\ref{finalsola}) is replaced by 
\be
\label{finalsolb}
\psi=\left(\begin{array}{c}1\\1\end{array}\right) r^{|k_1|} +  d e^{i\phi} \omega^{2\eta} \left(\begin{array}{c}1\\-1\end{array}\right)r^{-|k_1|}.
\ee
And for $k_1<0$,  $\eta>1/2$, eq.(\ref{finalsola2}) is replaced by 
\be
\label{finalsolb2}
\psi=\left(\begin{array}{c}1\\1\end{array}\right) r^{|k_1|} +  id' \omega^{2\eta} \left(\begin{array}{c}1\\-1\end{array}\right)r^{-|k_1|}.
\ee

It is now easy to follow the discussion in \S3.1 to calculate the  two-point function on the boundary in this case. 
A Fermi surface will  arise for $k_1=k_F$ if  the growing solution (this is the first term  on the RHS of 
 eq.(\ref{finalsola}), (\ref{finalsolb}) for this value of $k_1$ is purely normalisable in $AdS_4$.
The Green's function one gets by expanding around this value of momentum is 
\be
\label{tpt2}
G_{R22}={Z\over \omega- v_F(|k|-|k_F|)+iD e^{i\phi} \omega^{2\eta} }
\ee
where $Z,D$ are constants. For $\eta >1/2$, the phase $e^{i\phi}=1$, whereas for $\eta<1/2$ the phase is in general complex.  

This result is  very similar to what was obtained in the eRN case in \cite{SSL},\cite{MIT},\cite{CSZ},\cite{MIT2}. The result also 
agrees with the general considerations in \footnote{In fact eq.(\ref{tpt2}), for $\eta<1/2$, 
 is  of the scaling form proposed in \cite{Senthil} and satisfies the inequalities in eq.(9) and (18) of \cite{Senthil}.} \cite{Senthil}.

For $\eta>1/2$  there is a well-defined quasi-particle with a linear dispersion and  a  width which goes like 
$\omega^{2\eta}$. For $1/2<\eta<1$,   the width is broader than the Fermi liquid case. 
For $\eta<1/2$ the behaviour is more novel. The last term in the denominator going like $\omega^{2\eta}$ 
dominates both the real and imaginary parts of the $\omega$ dependence.  As a result there is no well-defined quasi-particle, since the residue vanishes at the pole. 
Finally for $\eta=1/2$, as discussed in Appendix D the Green's function actually needs to be modified and takes the form, 
\be
\label{mf}
G_{R22}={Z\over  v_F(|k|-|k_F|)+d_1 \omega\log{\omega} + d_2 \omega }
\ee
where $d_1$ is real and $d_2$ complex. 

Unlike the eRN case, when $\alpha=\delta$ and $\gamma\ne 1$, the geometry has no scaling symmetry. Despite this fact  
eq.(\ref{eomn}) has a scaling symmetry for all values of $\gamma$, when $\beta+\gamma=1$, under which $r\rightarrow \lambda r, 
\omega\rightarrow \lambda^{2\gamma-1} \omega$ with $k_1$ being invariant \footnote{In fact a suitable change of variables can map the 
eq.(\ref{eomn}) for all values of $\gamma$, and $\gamma+\beta=1$, to the case $\gamma=1,\beta=0$.}.
This scaling symmetry results in the  the complex part characterized by the exponent $\eta$ being of 
 power law type in the frequency. One difference is that in our case the mass and charge of
the bulk fermion do not enter in $\eta$ explicitly but only through    $k_F$, which does depend on these parameters. 

Before closing this subsection it is worth commenting  that while on general grounds we expect a value of $k_F$ to exist 
 for which the bulk solution is purely normalisable, leading to a singularity in $G_{R22}$,  we have not investigated this feature
in detail, e.g., in the solutions discussed in Appendix F which are asymptotically $AdS_4$. We leave such an analysis, 
along with the related calculation of 
$Z,v_F,D$ which appear in eq.(\ref{tpt2}), for the future. 

\subsection{Case $\beta + \gamma <$1 }
In this case the essential features of the solution can be deduced by setting the $k_1$ dependent terms in eq.(\ref{eomn}) to zero. 
This can be seen to be  self-consistently true. In fact the essential point was already made in \S3.2.
Setting the $k_1$ dependent term to vanish in eq.(\ref{eomn}) gives the solution eq.(\ref{nhpsi}). When eq.(\ref{psm})
 is met the solution reduces to (\ref{ppsi}). 
This happens before the $k_1$ term becomes important because 
\be
\label{relaa}
\omega^{1\over 2\gamma-1}\ll \omega^{1\over \gamma-\beta}
\ee
when $\beta+\gamma<1$ as discussed in \S3.2 around eq.(\ref{reginta}).

The reader might worry that this argument is a bit too quick. We will examine it more carefully in Appendix C
 and find that it is indeed justified. 
The more careful analysis shows that the region eq.(\ref{reginta}) where the solutions reduces to the form, 
eq.(\ref{ppsi}) should be thought of as being obtained by keeping  ${r / \omega^{1 /( 2\gamma-1)}}$ fixed and large while taking 
$\omega\rightarrow 0$.

Let us  now examine the consequences of eq.(\ref{ppsi}).  
Note in particular that at leading order  $y_+$ and $z_-'$ have a relative phase  which is imaginary.  
 
In the following discussion it  will be   useful at this stage to to define two basis vectors, 
\be
\label{basisvec}
\psi_+=  \left(\begin{array}{c} 1 \\0 \end{array}\right), \psi_-=\left(\begin{array}{c} 0 \\1 \end{array}\right)
\ee
and express the leading order  answer as 
\be
\label{fform}
\left(\begin{array}{c} y_+\\z_-' \end{array}\right)=\psi_+ -i \psi-
\ee

Starting from the region eq.(\ref{reginta})
 and going further towards the boundary the $k_1,m,A_t$ dependent terms we have been neglecting will come into play 
and the form of the solution
will deviate from eq.(\ref{fform}).  
Asymptotically, towards the boundary,  the solution obtained from both  $\psi_\pm $  will take the form given in eq.(\ref{formas}). 
Let $C_{1\pm}, C_{2\pm}$ be the values for the coefficients $C_{1,2}$ which appear in eq.(\ref{formas})
 when we start with $\psi_\pm$ and evolve the solution
towards the boundary respectively. 
Then, the net value of $C_1$ we get starting from eq.(\ref{fform}) is 
\be
\label{netform}
C_1=C_{1+}-iC_{1-}
\ee
Now notice that the equations eq.(\ref{yfirst}), eq.(\ref{zfirst}) are both real, therefore   $C_{1\pm}$ will be both real as well. 

As in the discussion for the $\beta+\gamma>1$ case a Fermi surface  arises at $k_1=k_F$ 
 when $C_1$ vanishes at this momentum
 as $\omega \rightarrow 0$. However,   since $C_{1+}$ and $C_{1-}$ are  real this actually  imposes two conditions
\be
\label{condfs}
C_{1+}=0 \,,\, C_{1-}=0
\ee
which must both be met by adjusting only one real variable - the momentum $k_1$. 
Generically, this will be impossible to do.

Our conclusions will be discussed more throughly in the following subsection. 
We will see that starting from $\beta+\gamma=1$ as we go into the region where $\beta+\gamma<1$, 
there is no locus 
  in momentum space  about which there are 
quasi-particle excitations with a width that vanishes as $\omega$ vanishes. However, for $\beta+\gamma<1$, but close to unity,
there is a surface about which the excitations have a frequency independent width (at small $\omega$) which is  much smaller
than the chemical potential and which vanishes as $\beta+\gamma\rightarrow 1$. 

Let us close by again mentioning that the  approximations made in the analysis above  of neglecting the $k_1$ dependent terms 
is  examined more carefully in Appendix C and found to be indeed valid.

\subsection{The Transition from $\beta+\gamma=1$ to  $\beta+\gamma<1$}
It is useful to discuss the transition from $\beta+\gamma=1$ to $\beta+\gamma<1$ in more detail. 

Let us start with the case $\beta+\gamma=1$ and 
 first consider  the case when the exponent $\eta$ in eq.(\ref{tpt2}) satisfies the condition $2\eta>1$. In this case, at small $\omega$,
\be
\label{formG}
G_{R22}= {Z\over \omega-v_F(|k|-k_F) + i d_1 \omega^{2\eta}},
\ee
and as was mentioned above  
there are well-defined quasi-particle excitations about  the Fermi surface.  

Suppose we now lower the value of $\beta+\gamma$  so that $\beta+\gamma=1-\epsilon, \epsilon \ll 1$.
The bulk fermion solution with momentum $k_F$ will not be purely normalisable any more and our arguments in the previous subsection 
 show that   the Green's function takes the form, 
\be
\label{formg2}
G_{R22}={Z\over \omega-v_F(|k|-k_F) + \Delta_1+i \Delta_2 + i d_2 \omega^{2\eta}}
\ee
where $\Delta_{1},\Delta_2$, are $\omega$ independent and  vanish when $\epsilon\rightarrow 0$. 
We see that $\Delta_1$ can be absorbed by a shift in \footnote{Alternatively, e.g., in the canonical ensemble, 
 we can absorb it into a shift in the chemical potential $\mu$.}  $k_F$, $k_F\rightarrow k_F -{\Delta_1\over v_F}$. About this new Fermi momentum we get,
\be
\label{formg33}
G_{R22}={Z\over \omega-v_F(|k|-k_F) +i \Delta_2 + i d_2 \omega^{2\eta}}
\ee
so that the excitations  have a width $\Delta_2$, which does not vanish as $|k|\rightarrow k_F,\omega\rightarrow 0$, 
and is therefore very broad.
In summary, the well-defined quasi-particle which existed at $\beta+\gamma=1$ has therefore disappeared  
 at  $\beta+\gamma<1$  
\footnote{The width  $\Delta_2$, while it does not vanish when $|k|\rightarrow k_F$,  is small compared to the chemical potential for  $\epsilon \ll 1$.}.

Next, let us turn to the case when the exponent $\eta$ in eq.(\ref{tpt2}) satisfies the condition $2\eta<1$. 
In this case there is no sharply defined quasi-particle  even when $\beta+\gamma=1$. 
We define, 
\be
\label{defkt}
k_T\equiv (|k|-k_F). 
\ee
Taking $|k_T|$ fixed and small compared to $k_F$, 
 and regarding $G_{R22}$ as a function of $\omega$ 
 there is a pole \cite{MIT4} at $\omega=\omega_*+i\Gamma$ with   
\be
\label{polea}
\omega_*\sim \Gamma \sim |k_T|^{1\over 2\eta}.
\ee
The pole has  vanishing residue, 
\be
\label{resvan}
Z_{res}\rightarrow 0,
\ee
and  is also broad, 
\be
\label{widthaa}
{\Gamma\over w_*} \rightarrow 1.
\ee
as $k_T\rightarrow 0$.

Let us now lower  $\beta+\gamma$ in this case to the value $\beta+\gamma=1-\epsilon$. This
  results\footnote{The term linear in $\omega$ in the denominator is dropped
compared to $\omega^{2\eta}$ at small $\omega$.} in  a Green's function,
\be
\label{formg3}
G_{R22}={Z\over -v_F(|k|-k_F) + \Delta_1+i \Delta_2 + D e^{i\phi} \omega^{2\eta}},
\ee
where $\Delta_{1,2}$ vanish as $\epsilon\rightarrow 0$. 
Shifting $k_F$ this can be written as
\be
\label{formg4}
G_{R22}={Z\over -v_F(|k|-k_F) + |\Delta| D  e^{i\phi} e^{-i\pi \eta} + D e^{i\phi} \omega^{2\eta}}
\ee
where $|\Delta|$ is determined by $\Delta_1,\Delta_2$ and the shift in $k_F$.
For $|k|\rightarrow k_F$ the pole in $\omega$ lies at 
\be
\label{pole}
\omega_*=-i|\Delta| 
\ee
This gives rise to a width which  does not vanish when $(k-k_F)\rightarrow 0$.

In summary we see that    when $\beta+ \gamma<1$ the   excitations become very broad and acquire a width which is non-vanishing at zero
frequency.
There is still a locus in momentum space, at $|k|=k_F$, which we can call the Fermi
surface,
 with the energy of the excitations, defined as the real part of  $\omega$,   extending down to zero energy as the momentum approaches $k_F$.
However, a more precise definition of the Fermi surface can be taken to be the locus where  Green's function with $\omega=0$ has a pole in momentum,
and across which it changes sign.  With this definition, there is no Fermi surface for $\beta+\gamma<1$,
since  the excitations have a non-zero
 width even at zero energy, as mentioned above \footnote{We thank Mohit Randeria for explaining this definition of the Fermi surface and for
related discussion.}.

\section{Conclusions}

The  gravity system  studied in this paper has a scalar, the dilaton, and two couplings $\alpha$ and $\delta$
which appear in the action given in eq.(\ref{dildepf}), (\ref{dildeppot}) and which 
determine how the dilaton enters in the gauge coupling and the  potential respectively. 
Instead of $\alpha,\delta$ it is sometimes more convenient to  use the parameters, $\beta, \gamma,$ which appear in the metric
eq.(\ref{ansatz2}) and  are given in terms of 
$\alpha,\delta,$ in eq.(\ref{case1}), (\ref{case11}).   

Extremal black branes in this system were studied in \cite{Kiritsis}.
Here we have  studied a charged fermion in the extremal black brane background  and calculated the two-point function for the corresponding
fermionic operator in the  dual strongly coupled field theory. The black brane background has rotational symmetry in the two spatial 
directions and the two-point function inherits this symmetry,   At small frequency, which is the focus of our investigation, the essential
features of this two -point function can be deduced from the near-horizon geometry of the extremal black hole. 

\subsection{Results}

Our results depend on the parameters $\beta,\gamma$, in particular on the combination \footnote{The distance to the horizon
for the variable $\zeta$, eq.(\ref{defzeta}),
 in terms of which the fermion equation of motion becomes of Schrodinger form is governed by $\beta+\gamma$.
It is infinite for $\beta+\gamma>1$,  logarithmically infinite for $\beta+\gamma=1$, and finite for $\beta+\gamma<1$.} $\beta+\gamma$. 
\smallskip
\noindent
$\bullet$
When $\beta + \gamma>1$ we find that close to the Fermi-surface there are well-defined quasi-particles, with a linear
(i.e. relativistic)  dispersion relation and a width which is exponentially suppressed in $\omega$. The precise form of the 
Green's function is given in eq.(\ref{ttp}). 

\smallskip
\noindent
$\bullet$
This behaviour undergoes a transition when $\beta+\gamma<1$. In this case   there are no  well-defined 
 quasi-particle excitations.   
 Instead the low-energy excitations become very broad with a  width
which does not vanish at small frequency. See the concluding paragraph of \S3.5 for  more discussion on the  Fermi surface. 

\smallskip
\noindent
$\bullet$
The transition region $\beta+\gamma=1$ is also very interesting. 
 In terms of the parameters $\alpha,\delta$ which appear in the 
Lagrangian for the system, this corresponds to two lines, $\alpha=\pm \delta$. 
The $\alpha=-\delta$ line corresponds to an extremal RN geometry. The fermionic two-point function in this case
is well studied and known to exhibit
 interesting  non-Fermi liquid behaviour. Here we focus on the other case,  the $\alpha=\delta$ line, for which 
 the extremal geometry has vanishing entropy and the near-horizon geometry has no scaling symmetry. 
Despite this difference 
we find that bulk fermion equation acquires a scaling symmetry analogous to that in the eRN case and 
the two-point function again exhibits non-Fermi liquid behaviour.
  The precise form of the Green's function is given in eq.(\ref{tpt2}) and depends on the parameter $\eta$. 
When $2\eta<1$, there are   no well-defined   quasi-particles excitations close to the Fermi surface. 
When $2\eta>1$, there are  well-defined  quasi-particle, with a width which vanishes as $\omega\rightarrow 0$,
although this width  can be much broader  than in Fermi-liquid theory. 
When $2\eta=1$, one gets a marginal Fermi liquid \footnote{One difference with the eRN case is that the mass and charge of the bulk fermion
enters in $\eta$ only through their dependence on the Fermi-momentum, $k_F$.}.
The quasiparticles, when they exist for the $\beta+\gamma=1$ case, get very broad when  $\beta+\gamma$ becomes less than unity. 

\smallskip
\noindent
$\bullet$
The transition between these behaviours occurs in a smooth way. More precisely,  the Green's function evolves in a smooth manner
as the parameter $\beta + \gamma$ is varied. The underlying reason for this is that the background geometry itself evolves smoothly.
\subsection{Discussion}

It is worth trying to phrase our results in terms of the semi-holographic description which was proposed in \cite{FP}. 
The near-horizon region of the geometry  corresponds to a  strongly coupled field theory sector in this description, which is coupled to
 bulk fermionic excitations localized away from the near-horizon region. The bulk fermionic excitations by themselves are weakly coupled
and form a sea which is essentially responsible for the Fermi surface in the boundary theory. 
The coupling between the two sectors allows the bulk fermions to decay and gives rise to their width. 
When $\beta+\gamma>1$ this decay width  is small, since it arises  due to tunneling through a WKB barrier. 
This  results in a  width  which is highly suppressed 
with an essential singularity as $\omega\rightarrow 0$. 
 As $\beta+\gamma \rightarrow 1$ the barrier is lowered and the bulk fermions can decay more easily into degrees of freedom in the 
 strongly coupled sector, 
 resulting in a decay width which is only power law suppressed. 
Finally, when $\beta+\gamma <1$  the decay  process  is sufficiently enhanced and leads to a 
 width which is non-vanishing even as $\omega\rightarrow 0$,
leaving  no sharply  defined quasi-particles   in the  excitation  spectrum. 

In fact in our   analysis we did not use the information about  the full geometry but only the geometry in the 
 near-horizon region. The full geometry depends on many more 
details of the model including the dependence, even  away from the run-away region where $\phi\rightarrow \pm \infty$, of the
 gauge coupling function and the potential on the dilaton. It is therefore less universal than the near-horizon geometry which  is in fact 
often  an attractor. 
 This is very much in the spirit of the semi-holographic description, in effect we only relied on the gravity dual for 
the strongly coupled sector, and did not use much information about the  gravity solution  away from the horizon since 
in the end that would have  
given rise to a weakly coupled bulk fermion whose dynamics can be understood in field theoretic terms anyways \footnote{
Some examples of gravity solutions which interpolate between $AdS_4$ and the solution eq.(\ref{ansatz2}) in the near-horizon region
are discussed in Appendix F. These are obtained 
 with reasonable potentials and gauge coupling functions.
It is worth studying these examples further  
to calculate the  value of  $k_F$ (for $\beta+\gamma\ge 1$) and the 
the value of the residue
and $v_F$ in eq.(\ref{tpt2}), (\ref{formg33}), (\ref{formg4}), in them.}.

The basic lesson then from this paper is that a range of interesting behaviours can arise by coupling fermions
to a strongly coupled sector with a gravitational dual of the kind considered here. 
This includes both Fermi liquid and  non-Fermi liquid behaviour, transitions 
between them, and transitions from a non-Fermi liquid state to one  where there are no well-defined  
    quasi-particles since the excitations  have  become very broad and essentially disappeared. 
 Moreover, this  can happen when the strongly coupled sector has reasonable thermodynamics behaviour consistent in particular
with the third law of thermodynamics, since the  gravity background has vanishing entropy at extremality.

An important feature about our system is that the  dramatic changes in the  behaviour of the fermionic Green's function which we have found
 are {\it not} accompanied by any phase transition or significant changes in the thermodynamics or 
transport properties.  
The entropy densty or  specific heat, for example, scale as   given by eq.(\ref{entropne})
 and smoothly changes as $\beta+\gamma$ is lowered from a value greater than unity to less 
than unity. Similarly, the DC or optical conductivity   also  changes smoothly, eq.(\ref{dccond}), eq.(\ref{opcond}), 
 eq.(\ref{relnu}). In fact  the background geometry itself changes smoothly, as was mentioned above, 
this is  the  root cause  for the  smooth behvaiour in transport and conductivity. 
On general grounds the gravity system should  correspond to a strongly coupled field theory in the large $N$ limit. In this limit there are 
many extra degrees of freedom besides the fermionic ones we have focussed on.  
And these extra degrees of freedom do not undergo any significant change in their properties even 
though the fermionic ones we have focussed on do,
resulting in the smooth changes in thermodynamics and transport.

The large $N$ limit is the price we pay for the having a tractable gravity description.
At finite $N$ one would expect that the transitions seen in the behaviour of the fermion correlator will also manifest itself
in phase transitions or big qualitative changes in thermodynamics and transport. Preliminary evidence for this is the fact that the conductivity
in our set up already has $1/N$ corrections  which see the changes in the nature of the fermion two-point function. 
This was investigated in \cite{MIT3}, \cite{MIT4}, where it was found that for a Green's function of the type in 
 eq.(\ref{tpt2}) there would be corrections to conductivity 
of the form
$\sigma \sim {1\over N} T^{-2\eta}$.  Since  we have found non-Fermi liquid behaviour  to arise from in 
a wide variety of  gravitational backgrounds
it is reasonable to hope that it will persist for some  finite $N$ strongly coupled  theories as well.

There are several directions for future work. It will be interesting to generalize the investigations of this paper 
to higher dimensions \footnote{See for example \cite{Chen:2010kn}.}. 

Going beyond effective field theory, it is important    to try and  embed the  class of  gravity systems studied  here in string/M theory. This would put constraints on allowed
values of $\alpha,\beta$ and also the charges and masses  for the Fermion fields which determine $k_F$ and the exponent 
$\eta$ in eq. (\ref{tpt2}). Allowed ranges of these parameters would then determine which kinds of non-Fermi liquid theories are 
theoretically speaking  allowed and when transitions of various kinds are allowed.  Embedding in string/M theory 
is also important for  deciding whether our approximation of classical two -derivative gravity is
a controlled one,  as was  
discussed in \S2.4.  For some progress towards providing such embeddings  see  \cite{GSW},  \cite{DGKV} 
\footnote{For example, our IR effective action (\ref{genact}) with parameter $\alpha = \sqrt{3}, \delta = - 1/\sqrt{3}, V_0 = - 12 \sqrt{3}$ can be obtained from M-theory on Sasaki-Einstein space from eq.(4.3) of \cite{GSW}, by setting $\chi = 0$ and $h \equiv \pm 1 \mp e^{4 \phi/\sqrt{3}}$, in the regime where $|e^{4 \phi/\sqrt{3}}| \ll 1$. However, the near horizon behaviour of the  numerically obtained solution in \cite{GSW} is
 different from our solution (\ref{case1}), (\ref{case11}). It is worth studying this point further, we thank J. Gauntlett, J. Sonner and 
 T. Wisemann for a discussion.}. 
It will also be useful to ask whether this analysis can be extended beyond the case where the gravity theory is analyzed in the two-derivative approximation for example in Vasiliev theory \cite{Vasiliev}. 

Another direction would be to couple charged matter and study superconducting instabilities \cite{IKNT} along the lines of \cite{Gubser1}.  Or to allow for 
a bulk Fermi sea in the near-horizon region and incorporate the changes this leads to \cite{strange}, \cite{Hartnoll1}, \cite{Verlinde} 
\footnote{We thank S. Minwalla and S. Hartnoll
for a discussion on this point.}. 

Investigating  transitions of the kind we have found  in the presence of a magnetic field would also be an interesting extension. 
In this context it would be natural to also include an axion in the bulk theory, \cite{GIKPTW}.

Finally, only a very small class of possible attractor geometries have been studied here \footnote{Some references pertaining to the attractor mechanism are, \cite{GIJT}, \cite{oldattractor1}, \cite{oldattractor2}, \cite{oldattractor3}, \cite{oldattractor4},  and more recently \cite{KKS}.}. There is clearly a vast zoo waiting to be explored
and the behaviour of fermions in these additional backgrounds might hold even more surprises.

We leave these directions for the future.


\bigskip
\centerline{\bf{Acknowledgements}}
\medskip

We thank   S. Hartnoll,  G. Mandal,
S. Minwalla, S. Sachdev,  T. Senthil, R. Shankar, V. Tripathy,   S. Wadia and especially,  K. Damle, S. Kachru and M. Randeria,  
 for helpful discussions, and thank S. Prakash for  collaboration at early stages of this work.
SPT has benefitted greatly from two  Chandrasekhar Discussions meetings organized by the ICTS, on ``Applied String Theory'' and ``Strongly Correlated Systems and AdS/CFT'' and 
thanks the lecturers and participants of these meetings. He also thanks the organizers and participants of the Aspen Winter Conference on 
``Strongly Correlated Systems and Gauge Gravity Duality''. 
NK, PN and SPT acknowledge funding from the Government of India,  
and thank the people of India for  generously supporting research in string theory. 

NI would like to dedicate this work to the victims of the 2011 Tohoku Earthquake and Tsunami in Japan.

\newpage

\appendix
\section{Other Near-Horizon Geometries.} 
In this appendix we discuss another kind of near-horizon extremal solution which is obtained  from the equations of motion and is also 
consistent with the ansatz, eq.(\ref{ansatz2}). 
The essential idea is to meet eq.(\ref{em2}) and eq.(\ref{em1}) as in eq.(\ref{condk}), eq.(\ref{cond1ca}), 
eq.(\ref{condgamma}). 
However, eq.(\ref{em3}), eq.(\ref{em4}) are now met by assuming that 
\be
\label{condappb}
 2 \beta <  -k (\alpha + \delta)
\ee
so that the LHS in eq.(\ref{rel3}), eq.(\ref{rel4})  is more important at small $r$ than the RHS. 
This yields the equations, 
\begin{eqnarray}
k^2 &=& \beta (1 - \beta) \\
C_a^2 (1 + \beta + k \delta) (1 + 2 \beta + 2 k \delta) &=& - V_0 \\
\gamma -1 &=& \delta k \\
2 C_a^2 k (1 + 2 \beta + 2 k \delta) &=& \delta V_0 \\
C_a^2 \beta (1 + 2 \beta + 2 k \delta ) &=& - {1 \over 2} V_0 
\end{eqnarray}
The solution is
\begin{eqnarray}\label{case2}
\beta = {1 \over 1 + \delta^2}  \hspace{5mm} 
\gamma = {1 \over 1 + \delta^2}  
 \hspace{5mm} 
k = - {\delta \over 1 + \delta^2}  
 \hspace{5mm} 
C_a^2 = - V_0 \ {(1 + \delta^2)^2 \over 2 (3 - \delta^2)} 
\end{eqnarray}
This is valid for those values of  $(\alpha,\delta) $  which satisfy the conditions, 
\begin{enumerate}
 \item $2  <  \delta (\alpha + \delta)$ and
 \item $\delta^2<3$. 
\end{enumerate}
The first of these conditions, arises from eq.(\ref{condappb}). 

Let us  describe the solution discussed in \S2 as Case(1), and the solution discussed in this appendix as Case(2). 
For a fixed $\alpha$ ($> 0$) and $\delta$ small, we are in  region where Case(1) is valid.
 If we then increase $\delta$, this solution
 breaks down when $Q_e^2$ turns negative (i.e $\delta (\alpha+\delta)=2$), and we enter into the region where Case(2) becomes valid.

\section{More on Conductivity}
Here we provide some details for the calculation of the conductivity. 
The metric in the near horizon extremal  geometry 
is 
\begin{equation}
\label{metric}
 ds^2 = - C_a^2 r^{2 \gamma} dt^2 + {dr^2 \over C_a^2 r^{2 \gamma}} + r^{2 \beta} (dx^2 + dy^2)
\end{equation}
with $e^{\phi} = r^{k}$ and  the gauge field being
\begin{equation} \label{gaugefieldsoln}
 F={Q \over r^{2 \alpha k + 2 \beta}} dt \wedge dr.
\end{equation}
The constants, $\gamma,\beta,C_a^2,Q^2$,  are given in eq.(\ref{case1}) and (\ref{case11}).

Comparing with \cite{GKPT} eq.(3.3) 
\begin{equation}
 ds^2 = - g(\hat{r}) e^{- \chi(r)} dt^2 + {d\hat{r}^2 \over g(\hat{r})} + \hat{r}^2 (dx^2 + dy^2),
\end{equation}
 we have, 
\begin{eqnarray}
 \hat{r} &=& r^{\beta} \\
g(\hat{r}) &=& C_a^2 \beta^2 \hat{r}^{{2 \over \beta}(\beta + \gamma -1)} \\
e^{\chi(\hat{r})} &=& \beta^2 \hat{r}^{2(1- {1 \over \beta})}
\end{eqnarray}

The Schroedinger variable defined in eq.(3.8) of \cite{GKPT}, also eq.(\ref{defz}) above, is then, 
\begin{equation}
 z = -{1 \over (2 \gamma -1) C_a^2} {1 \over r^{2 \gamma -1}} = -{1 \over (2 \gamma -1) C_a^2} {1 \over \hat{r}^{2 \gamma -1 \over \beta}}
\end{equation}
The horizon lies at $r \ra 0$ or $z \ra -\infty$. The Schroedinger potential is  as given in eq.(3.15) of \cite{GKPT}
\begin{equation}
 V(z) = \left( {f'' \over f} + 16  {g e^{-\chi} \over f^2} {Q^2 \over r^4} \right)
\end{equation}
where $f = 2 e^{\alpha \phi}$. Note the factor of 16 which was not there in \cite{GKPT} eq.(3.15). This is because their convention for $Q$ differs from our $Q$ by a factor of $4$. This can be seen by comparing eq.(3.14) of \cite{GKPT} with eq.(\ref{gaugefieldsoln}).
 Plugging in the values of $k,\gamma,C_a,Q$ from eq.(\ref{case1}) and eq.(\ref{case11}), we get
\begin{equation}
 V = {2 \over z^2} {(4 + \alpha^2 -\delta^2) \left(4 +( \alpha- 2 \delta)(\alpha + \delta)\right) 
\over \left( 4 + (\alpha - 3 \delta) (\alpha + \delta) \right)^2 },
\end{equation}
leading to $c$ as given in eq.(\ref{valc}) above.

\section{Appendix C}

Most of this appendix will deal with  a more careful analysis justifying our neglect of the $k_1$  dependent terms in eq.(\ref{eomn}) in  \S3.4. 
However, before we embark on that discussion let us note the following. 
We had argued  after eq.(\ref{condfs}) that generically the two conditions
 required for the existence of a Fermi surface will not be met by tuning just 
$k_1$  which is one real variable. 
However, one might have a lingering doubt as to whether this non-genericity is  built into the very nature of  
  eq.(\ref{yfirst}), (\ref{zfirst}) and follows perhaps from its symmetries. We will now set this doubt to rest. 

The equations eq.(\ref{yfirst}), (\ref{zfirst}) are invariant  under an exchange of $y_+\leftrightarrow z_-'$ accompanied
by a simultaneous change in the parameters, $(m,\omega,A_t)\rightarrow -(m,\omega,A_t)$ with $k_1$ being kept the same. 
The wave functions $\psi_+$ and $\psi_-$, eq.(\ref{basisvec}), are exchanged under this transformation. 
This means 
\be
\label{relca}
C_{1+}(m,\omega,q,k_1)=C_{2-}(-m,-\omega,-q,k_1)
\ee
and 
\be
\label{relcb}
C_{1-}(m,\omega,q,k_1)=C_{2+}(-m,-\omega,-q,k_1)
\ee
In turn, this means that the two conditions, eq.(\ref{condfs})
 can be stated in terms of the coefficients $C_{1+}, C_{2+}$ that characterize the 
asymptotic behaviour of just one of the solutions $\psi_+$, as 
\be
\label{condsappc}
C_{1+}(m,\omega,q,k_1)=0 \,,\hspace{1mm}\, C_{2+}(-m,-\omega,-q,k_1)=0 \,.
\ee
It is now clear that both of these will not be generically satisfied for any $k_1$.

\subsection{Neglecting the $k_1$ Dependent Terms}
It is well known that perturbation theory about an extremal background is often fought with subtleties due to the 
extremal nature of the horizon. For this reason we would like to be carry out a more careful analysis that justifies
 our approximation of neglecting the $k_1$ dependent terms in eq.(\ref{eomn}) when $r$ ranges from being very close to the horizon
to values of order eq.(\ref{reginta}) where the form of $\psi$ in eq.(\ref{ppsi}) can then be used to extract the leading $\omega$ dependence 
as  explained in \S3.4.

We  use bra-ket notation below and denote, 
\be
\label{defnewpsi}
 |\psi> \equiv (\begin{array}{c}  y_+\\z_-'\end{array})
\ee
The unit norm eigenstates of $\sigma_2$ with eigenvalue $\pm1$ will be denotes by $|\pm>$ respectively below. 

The equations eq.(\ref{yfirst}), eq.(\ref{zfirst}) are
\be
\label{braceq}
r^{\beta+\gamma}\partial_r|\psi>=-k_1\sigma_1|\psi>+i\sigma_2 r^{\beta-\gamma}\omega|\psi>
\ee
Neglecting the $k_1$ dependent term gives the zeroth order solution eq.(\ref{nhpsi}),
\be
\label{defpsi0} 
|\psi_0>=e^{-i\omega z} |->.
\ee

Let 
\be
|\psi_1> = F_1 |+>
\ee
 be the first order correction induced by the $k_1$  dependent term we have neglected.
We get from eq.(\ref{braceq})
\be
\label{pertt}
\partial_z F_1-i \omega F_1= -k_1 r^{\gamma -\beta}<+|\sigma_1|-> e^{-i\omega z}
\ee
where $z$ is given in terms of $r$ in eq.(\ref{defzvch}). 
This gives,
\be
\label{pertth}
\partial_z F_1-i \omega F_1=ik_1 r^{\gamma -\beta} e^{-i\omega z}
\ee

The solution is 
\be
\label{sola}
F_1(z) =e^{i\omega z} i d_2 k_1 \int_{-\infty}^z dy    y^{{\gamma-\beta \over 1-2\gamma} } e^{-2i\omega y} 
\ee
The constant of integration is chosen by the fact that the full solution must be purely in-going at the horizon. 
The coefficient $d_2$ is determined by $\gamma,\beta$. 
Now notice that  since $\beta+\gamma<1$ the second integral is convergent. 
By rescaling variables, $x=\omega y$ we get
\be
\label{solb}
F_1(z)=e^{i\omega z} i d_2 k_1  \omega^{{1-\gamma-\beta \over 2\gamma-1}}  \int_{-\infty}^{\omega z}  dx
 x^{{\gamma-\beta \over 1-2\gamma}} e^{-2ix}.    
\ee

We need the correction to be small in the region eq.(\ref{reginta}) so that eq.(\ref{ppsi}) is a good 
approximation. For $\beta+\gamma<1$ the power ${\gamma-\beta \over 1-2\gamma}<-1$, so that the integral is 
convergent at the lower end, $-\infty$, but could diverge at the upper end if $\omega z \rightarrow 0$. 
This tells us that to keep the effects of the $k_1$ dependent term small we should work in the region where 
$|\omega z|\ll 1$ but is kept fixed as $\omega \rightarrow 0$. In terms of $r$ this becomes the condition that 
$({r / \omega^{ 1\over 2\gamma-1}})\gg 1$ and kept  fixed as $\omega \rightarrow 0$. 
In this region the upper limit of the intergal in eq.(\ref{solb}) is a fixed number and the integral converges. 

It then follows that the correction $F_1(z)$ is small for $\omega \rightarrow 0$, compared to the leading term which is order unity, 
since it is suppressed by $\omega^{{1-\gamma-\beta \over 2\gamma-1}}$, which is small, due to  $\beta+\gamma < 1$ and also $2\gamma-1>0$ from eq. (\ref{thermoconstraint}). 

Thus we see that a more careful analysis, where we have calculated the first corrections, shows that the $k_1$  dependent
terms can indeed be neglected in our analysis of section \S3.4. 

Before concluding let us make two comments. 
First, a similar analysis could have been repeated keeping the $m_1$ dependent terms in the equation of motion and shows 
that these are  even less worrisome that the $k_1$ dependent ones, as we expect on the basis of our arguments in \S3.2. 
Second,  note that when $\beta+\gamma=1$ the correction eq.(\ref{solb}) is logarithmically divergent at the horizon 
 (as $y\rightarrow -\infty$ in eq.(\ref{sola})).
This means the effects of the $k_1$ dependent terms  will not be small when $\beta+\gamma=1$  and we will need to include them 
in our analysis on par with the $\omega$ dependent terms.
 
\section{More on the $\beta+\gamma=1$ Case}
As mentioned in the \S 3.3 in this Appendix we discuss following issues.
\begin{itemize}
\item The detailed analysis to get the solution as in eq.(\ref{finalsola}), from the eq.(\ref{solxim}) and (\ref{solxip}), by series expanding the Hankel solution in eq.(\ref{solxim}) for small argument.
\item The analysis to reach eq.(\ref{finalsolb}) and (\ref{finalsolb2}) for the case $k_1<0$.
\item The marginal Fermi-liquid that arises for $k_1 = {1 \over 2}  (2 \gamma -1)$.
\end{itemize}

\subsection{Analysis with $k_1>0$}
Eq.(\ref{solxim}), for $z\ll1$, can be power law expanded to get a series expansion of $\chi_-$ and using eq.(\ref{solxip}) $\chi_+$ is,
\begin{eqnarray}\label{apndxd1}
\chi_- &=&  \tz^{\eta + 1} \left(c_1  + O(\tz^2)\right)  + i \left\lbrace \tz^{\eta +1} \left(d_1 + O(\tz^2)\right) +  \tz^{-\eta} \left (d_3 + d_4 \tz^2+ O(\tz^3)\right) \right\rbrace\\
\label{apndxd2}
\chi_+ &=& - \tz^{\eta } \left(c'_1  + O(\tz^2)\right)  - i \left\lbrace \tz^{\eta} \left(d'_1 + O(\tz^2)\right) +    \tz^{-\eta-1} \left( d'_4 \tz^2 + O(\tz^3) \right) \right\rbrace
\end{eqnarray}
where $c_1,d_1,d_3,d_4,c'_1,d'_1,d'_4$ are all constants depending on $\eta$. For small $\tz$, it is evident that $\chi_-$ dominates over $\chi_+$, (i.e. $i \tz^{-\eta} d_3$ term from $\chi_-$). Further calculations are divided into two cases depending on $\eta $,

\begin{itemize}

 \item {$0<\eta<{1 \over 2}$}.\\
In subleading order, $d_1$ terms dominate over $d_4$ term in $\chi_-$. So we get
\begin{eqnarray}
\chi_- &=&  i \tz^{-\eta} d_3 + \tz^{\eta+1} (c_1 + i d_1) \\
\chi_+ &=& - \tz^{\eta} (c'_1 + i d'_1)
\end{eqnarray}
Using eq.(\ref{defxipm}) and using the fact that $\tz^\eta  \gg \tz^{\eta +1}$, we get the original wave functions $y_+,z'_-$ as,
\begin{eqnarray}
y_+ &=& i \tz^{- \eta} {d_3 \over 2} - \tz^\eta {(c'_1 + i d'_1)\over 2}\\
z'_- &=& - i \tz^{-\eta} {d_3 \over 2} -\tz^\eta {(c'_1 + i d'_1) \over 2}.
\end{eqnarray}
Using eq.(\ref{deftz}) we can extract the $\omega$ dependence from $\tz$ in the above eq, and up to an overall $\omega$ dependent factor, which is not important, the complete wave function in terms of radial variable r can be written as,
\begin{equation}
\psi=\left(\begin{array}{c} y_+ \\ z_-' \end{array}\right)=\left(\begin{array}{c}1\\-1\end{array}\right) r^{k_1} + d  e^{i\phi} \omega^{2\eta} \left(\begin{array}{c}1\\1\end{array}\right)r^{-k_1}.
\end{equation}
where $d$ is real number dependent on $k_1$. This is the same as eq.(\ref{finalsola}).

\item {$\eta>{1 \over 2}$}.\\
In the subleading order $d_4$ term dominates over $d_1$ terms in $\chi_-$. So we get
\begin{eqnarray}
\chi_- &=&  i \tz^{-\eta} d_3 + i \tz^{-\eta+2} d_4 + c_1 \tz^{\eta+1}\\
\chi_+ &=& - i \tz^{-\eta+1} d'_4 - \tz^{\eta} c'_1
\end{eqnarray}
Using eq.(\ref{defxipm}) to go to variables $y_+$ and $z'_-$, and upon using the fact that $\tz^{\eta} \gg \tz^{\eta+1}$ and $\tz^{-\eta+1} \gg \tz^{-\eta+2}$ we get 
\footnote{We drop the terms proportional to $d_4'$ and  $d_4$  because they are purely imaginary and make a contribution suppressed by
powers of $\omega$ and $\omega^2$ respectively compared to the leading terms proportional to $d_3$ which is also purely imaginary. 
Note that on the other hand, a term proportional to 
$c_1'$ are purely real, so we keep this term.}
\begin{eqnarray}
 y_+ &=& i \tz^{- \eta} {d_3 \over 2}  -\tz^\eta {c'_1\over 2}\\
z'_- &=& - i \tz^{-\eta} {d_3\over 2}  -\tz^\eta {c'_1\over 2}
\end{eqnarray}

Again using eq.(\ref{defxipm}), eq.(\ref{deftz}), and up to some overall normalization the solution as,
\begin{equation}
\psi=\left(\begin{array}{c} y_+ \\ z_-' \end{array}\right)=\left(\begin{array}{c}1\\-1\end{array}\right) r^{k_1} 
  + i  d'  \omega^{2\eta} \left(\begin{array}{c}1\\1\end{array}\right)r^{-k_1}.
\end{equation}
where $d'$ is a  real number dependent on $k_1$. This is same as 
 eq.(\ref{finalsola2}).
\end{itemize}
\subsection{Analysis with $k_1<0$}

When $k_1<0$, the differential equations instead of eq.(\ref{chieqn1}) and (\ref{chieqn2}) will be now given by
\begin{eqnarray}
\label{chieqn1mk}
 (\tz \partial_z + \eta) \chi_+ &=& \tz \chi_- \\
\label{chieqn2mk}
(\tz \partial_z - \eta) \chi_- &=& - \tz \chi_+
\end{eqnarray}
Solutions to the above equations can be easily got by observing the following: $\chi_+,\chi_-$ now satisfy the same differential equation as $\chi_-,-\chi_+$ satisfied in eq.(\ref{chieqn1}) and (\ref{chieqn2}). Using this fact and the explicit solution from eq.(\ref{solxim}) we can write the solution as,
\begin{eqnarray}
\chi_+ &=& \sqrt{\tz} H^{(1)}_{{1 \over 2}  + \eta } (\tz)\\
\chi_- &=&  \tz^{-\eta} \partial_\tz (\tz^{\eta} \chi_+).
\end{eqnarray}
A similar analysis as before goes through with now $\chi_+$ being  the dominating solution. Again the analysis is divided into cases,

\begin{itemize}

\item {$0<\eta<{1 \over 2} $}.
\begin{eqnarray}
y_+ &=& i \tz^{- \eta} {d_3 \over 2} + \tz^{\eta} {(c'_1 + i d'_1)\over 2}\\
z'_- &=& i \tz^{-\eta} {d_3 \over 2} -\tz^{\eta} {(c'_1 + i d'_1)\over 2}
\end{eqnarray}
which gives the wavefunction in the radial variable $r$ given in eq.(\ref{deftz}) as,
\begin{equation}
\psi=\left(\begin{array}{c} y_+ \\ z_-' \end{array}\right)=\left(\begin{array}{c}1\\1\end{array}\right) r^{|k_1|} +  d    e^{i\phi} \omega^{2\eta} \left(\begin{array}{c}1\\-1\end{array}\right)r^{-|k_1|}.
\end{equation}
where $d$ is $k$ dependent real number. This is the same as eq.(\ref{finalsolb}).

\item {$\eta>{1 \over 2}$}.
\begin{eqnarray}
 y_+ &=& i \tz^{- \eta} {d_3\over 2} +\tz^{\eta} {c'_1\over 2}\\
z'_- &=& i \tz^{-\eta} {d_3\over 2} -\tz^{\eta} {c'_1\over 2}
\end{eqnarray}
which gives the wavefunction in the radial variable $r$ given in eq.(\ref{deftz}) as,
\begin{equation}
\psi=\left(\begin{array}{c} y_+ \\ z_-' \end{array}\right)=\left(\begin{array}{c}1\\1\end{array}\right) r^{|k_1|} + i d'  \omega^{2 \eta} \left(\begin{array}{c}1\\-1\end{array}\right)r^{-|k_1|}.
\end{equation}
where $d'$ is real number dependent on $k_1$.Therefore to leading order we get eq.(\ref{finalsolb2}). 
\end{itemize}

\subsection{Marginal Fermi Liquid for $k_1 = {1 \over 2}  (2 \gamma -1)$}
For the specific value of  $k_1 = {1 \over 2}  (2 \gamma -1)$, {\it i.e.}, $\eta = {1 \over 2}$, something interesting happens. We rework the above calculation with $\eta={1 \over 2}$ now, then eq.(\ref{solxim}) and (\ref{solxip}) become,
\begin{eqnarray}
\chi_- &=& \sqrt{\tz} H^{(1)}_1 (\tz) \,\\
\chi_+ &=& - \tz^{-1 \over 2} \partial_\tz (\sqrt{\tz} \chi_-) 
\end{eqnarray}
Expansion of Hankel function with index 1 for small $\tz$ will have a $\log$ term in the imaginary part in subleading order. More explicitly,
\begin{eqnarray}
\chi_- &=& {\tz^{3 \over 2}  \over 2} \left( 1+O(\tz^2) \right) + i \left\lbrace {c_1 \over \sqrt{\tz} }+ c_2 \tz^{3 \over 2} \log \tz + c_3 \tz^{3 \over 2} + O(\tz^2)  \right\rbrace\\
\chi_+ &=& -\sqrt{\tz} + i \sqrt{\tz} \left[ c_2' \log \tz  + c_3' \right]
\end{eqnarray}
$c_1,c_2,c_3,c'_2,c'_3$ are constants depending on $k_1$. Since $\tz \ll 1$ , $c_1$ term dominates over $c_2,c_3$ terms in $\chi_-$, and  $c'_2$ term dominates over $c'_3$ term in $\chi_+$. Note that $\chi_-$ has the most dominant term ${1 \over \sqrt{\tz}}$. Now using eq.(\ref{defxipm}) we get $y_+$ and $z'_-$ as,
\begin{eqnarray}
 y_+ &=& i  \left[ {c_1 \over \sqrt{\tz}}  + c_2' \sqrt{\tz} \log{\tz}  \right]  -\sqrt{\tz}  \\
 z'_- &=& - i  \left[ {c_1 \over \sqrt{\tz}} -c_2'  \sqrt{\tz} \log{\tz}  \right]  - \sqrt{\tz}
\end{eqnarray}
where we have used $\sqrt{\tz} \gg \tz^{3 \over 2}$. In terms of radial variable $r$, given in eq.(\ref{deftz}), and upto overall $\omega$ dependent factor,we get,
\begin{equation}
\psi=\left(\begin{array}{c} y_+ \\ z_-' \end{array}\right)=\left(\begin{array}{c}1\\-1\end{array}\right) r^{-{2 \gamma -1 \over 2}} +g_1 \omega \log \omega \ \left(\begin{array}{c}1\\1\end{array}\right)r^{2\gamma-1 \over 2}  + i d'  \omega \left(\begin{array}{c}1\\1\end{array}\right)r^{2\gamma-1\over 2}.
\end{equation}
where $g_1,d'$ are real numbers dependent on $k_1$. Using this wavefunction one gets the Green's function as written down in eq.(\ref{mf}).

\section{Scalar Two Point Function }
In this appendix we study a bulk charged massive scalar (denoted by $\Phi$, so that it is not confused with the dilaton) 
and calculate the   two-point function  for the corresponding operator
in the boundary theory. 

The action for the charged scalar is
\begin{equation}
 S=-\int d^4 x \sqrt{-g} \left((D_\mu \Phi)^2+m^2 \Phi^2 \right),
\end{equation}
where $D_\mu \Phi=(\partial_\mu - i q  A_\mu)\Phi$.

The equation of motion for the scalar field $\Phi = \Phi(r) e^{i \omega t + i \vec{k} . \vec{x}}$ is
\begin{equation}\label{eomscalar1}
  { 1 \over \sqrt{g}} \partial_r (\sqrt{g} g^{rr} \partial_r \phi) - (g^{ii} (k^2 - u^2) + m^2) \Phi = 0 
\end{equation}
where $u$ is as given in eq.(\ref{defu}). 
In extracting the correlator in the field  theory it is convenient to assume that there is a gravitational background which 
 extends  to $AdS_4$, with radius $L$,   on the boundary  and then use the AdS/CFT dictionary to calculate the boundary correlator. 
Asymptotically towards the boundary, 
the scalar field will behave as  
\begin{equation}\label{asymsoln}
 \Phi = C_1 r^{\Delta} + C_2 r^{- \Delta}
\end{equation}
where $\Delta$ is the positive root of the equation $\Delta (\Delta + 3) =  m^2 L^2 $. Then the two-point function on the boundary\footnote{Here we assume the standard quantization for this field.} for operator dual to $\Phi$ is,
\begin{equation}\label{prescrgreen}
 G_R = {C_2 \over C_1}
\end{equation}

To obtain the retarded Green's function, we impose ingoing boundary conditions at  the horizon. The near-horizon background  solution 
is given by eq.(\ref{case1}, \ref{case11}).  Setting $C_a$ to unity by rescaling $\vec{k},m,q$, as for the fermionic case gives, 
\begin{equation}\label{theeqn1}
 {1 \over r^{2 \beta} } \partial_r\left({ r^{2 \beta + 2 \gamma} \partial_r \Phi }\right)+ {(\omega + q A_t)^2 \over  r^{2 \gamma}} \Phi -
 { k^2 \over r^{2 \beta}} \Phi -m^2 \Phi = 0.
\end{equation}
In the near horizon region, the mass term is subdominant, since $r \ll 1$, and also the $A_t$ term can be neglected compared to
 the $\omega$ dependent for the same reason as in the fermionic case: the frequency dependence at small $\omega$ can be 
 extracted before the $A_t$ term becomes important. 
This gives, 
\begin{equation}
\label{theeqn}
{1 \over r^{2 \beta} } \partial_r\left({ r^{2 \beta + 2 \gamma} \partial_r \Phi }\right)+ {\omega^2 \over  r^{2 \gamma}} \Phi - 
 { k^2 \over r^{2 \beta}} \Phi  = 0.
\end{equation}

Just as in the fermionic case it is convenient to divide the analysis into three cases. 
\subsection{$\beta+\gamma>1$}
Here the equation can be analyzed in the WKB approximation.  
The analysis is similar to \S3.1 of the fermionic case and we will skip some details. 
It is easy to see that the  WKB approximation is self-consistently valid. Changing to the variable $\zeta$, eq.(\ref{defzeta}) gives 
that $\Phi$ satisfies the same 
equation as $y_+$, eq.(\ref{sea}). 
The resulting solution in the  region eq.(\ref{sr}) is 
\be
\label{solsa}
\Phi={A\over \sqrt{\hat{k}}}[f_++{i\over 2} e^{-2I} f_-]
\ee
We see that $f_-$ has a coefficient which is exponentially suppressed. The factor  $I$ in the exponential suppression is 
\be
\label{appvalI}
I=c_1 \bigl({|k|^{2\gamma-1}\over \omega^{(\beta+\gamma-1)}}\bigr)^{1\over \gamma-\beta}
\ee
where $c_1$ is given in eq.(\ref{valc1}).

Now suppose  the coefficients $C_1, C_2$ , eq.(\ref{asymsoln}), that $f_+$ gives rise to are $C_{1+}, C_{2+}$, and similarly for $f_-$. 
And suppose an expansion of these coefficients in a power series in $\omega$ has  leading order terms
denoted by $C_{1\pm}^{(0)}, C_{2\pm}^{(0)}$, the first corrections by $C_{1\pm}^{(1)},C_{2\pm}^{(1)}$ etc.  
Then the two-point function is  \cite{MIT2}
\be
\label{tpa}
G_R={C_{2+}^{(0)}+\omega  C_{2+}^{(1)}\omega + O(\omega^2) + {i\over 2} e^{-2I}[ C_{2-}^{(0)}+C_{2-}^{(1)}\omega + O(\omega^2)]\over 
     C_{1+}^{(0)}+C_{1+}^{(1)} \omega + O(\omega^2)+{i\over 2} e^{-2I}[C_{1-}^{(0)}+C_{1-}^{(1)}\omega + O(\omega^2)]}.
\ee

Physically, the coupling of the scalar 
to the near-horizon region is highly suppressed, as a result the imaginary part of the Green's function, which arises because
the scalar can fall into the black hole, is also exponentially small. 

\subsection{$\beta+\gamma<1$}

In this case the leading small frequency dependence can be obtained by studying the region, 
$1\gg ({\omega\over k})^{1\over \gamma-\beta}\gg r$, thus the $k_1$ dependent term in eq.(\ref{theeqn}) can be dropped, giving
\begin{equation}\label{scaleqregion1}
  r^{-2 \beta} \partial_r \left(  r^{2 \beta + 2 \gamma} \partial_r \Phi \right) + {\omega^2 \over  r^{2 \gamma} }\Phi = 0
\end{equation}.
Now define
\begin{equation}\label{variable}
 z = {\omega \over 2 \gamma -1}\  {1 \over r^{2 \gamma -1}}.
\end{equation}
Also define 
\be
\label{appeta}
\eta = {2  \beta \over 2 \gamma -1}.
\ee
 Then in terms of $z$ the differential equation eq.(\ref{scaleqregion1}) is
\begin{equation}
 z^{\eta} \partial_z \left( z^{-\eta} \partial_z \Phi\right) + \Phi= 0
\end{equation}
whose solution is
\begin{equation}\label{Phi}
 \Phi = z^{1 + \eta \over 2} H^{(1)}_{1 + \eta \over 2} (z).
\end{equation}
For $\beta+\gamma<1$ we can be in the region where   $z \ll 1$ consistent with the requirement 
$({\omega\over k})^{1\over \gamma-\beta}\gg r$ needed to drop the $k_1$ dependent term in eq.(\ref{theeqn}), 
since, for  $\omega \rightarrow 0$, 
 $\left( {\omega \over k} \right)^{1 \over \gamma - \beta} \gg r \gg \omega^{1 \over 2 \gamma -1}$. In this region, 
the argument of the Hankel function  is small and we get, 
\begin{equation}
\Phi =   d_1 + i d_2 \left({\omega \over r^{2\gamma-1}}\right)^{1+\eta},
\end{equation}
where $d_1,d_2$ are real and we have used eq.(\ref{variable}). 
In fact a constant and $1/r^{2\gamma-1+2\beta}=1/r^{(2\gamma-1)(1+\eta)}$ 
are the two independent solutions to  eq.(\ref{scaleqregion1}) in this region where
the $\omega$ dependent term can be neglected.   Denoting the constant solution as $\phi_+$ and the real solution going like 
$1/r^{2\gamma-1+2 \beta}$ as $\phi_-$ and taking the coefficients $C_{1\pm}, C_{2\pm}$ to be defined as in the previous case we get  
\begin{equation}
G = {C_{2+}+ i \ C_{2-} \ \omega^{1+\eta} \over C_{1+} \ + i \ C_{1-} \ \omega^{1+\eta}}.
\end{equation}
The coefficients $C_{2\pm}, C_{1\pm}$ in turn can be power series expanded in $\omega$ as in the previous subsection.

We see that in this case the imaginary part of $G$ is only power-law suppressed in $\omega$ with a power, $\eta$, eq.(\ref{appeta}),
 which is independent
of the momentum $k$ and the charge and mass  of the scalar. 

\subsection{$\beta+\gamma=1$}
In this case we define 
\be
\label{appdefphi}
\Phi = { \xi \over \sqrt{r}},
\ee
then in terms of $z$, eq.(\ref{variable}), eq.(\ref{theeqn})  becomes, 
\begin{equation}
\label{abeq} 
 z \partial_z (z \partial_z \xi ) + (z^2 - \nu^2) \xi = 0
\end{equation}
where 
\be
\label{defnuapp}
\nu^2 = {k^2 + {1\over4} \over (2 \gamma -1)^2}.
\ee
Eq.(\ref{abeq})   is a Bessel equation whose solution with ingoing boundary conditions  is
\begin{equation}
 \xi = H^{(1)}_{\nu} (z).
\end{equation}
When $z \ll 1$, the Hankel function can be power series expanded to get, upto an overall power of $\omega$, 
\begin{equation}
\label{asstp}
\Phi={1\over \sqrt{r}}[d_1 r^{\nu(2\gamma-1)} + i d_2 \omega^{2\nu} {1\over r^{\nu(2\gamma-1)}}].
\end{equation}
Denoting $\phi_{\pm} $ as the two independent real  solutions to the scalar equation in this region which appear 
in the   first and second terms on  the RHS  respectively of eq.(\ref{asstp}) and defining $C_{1\pm}, C_{2\pm}$ as  the coefficients of the non-normalisable and normalisable terms these give rise to, as above, we then get
the boundary two-point function to be, 
\be
\label{btwp}
G={C_{2+}+i C_{2-} \omega^{2\nu} \over C_{1+}+iC_{1-}\omega^{2\nu}}.
\ee
Note that in this case the imaginary part is again power-law suppressed but this time the power, $\nu$, eq.(\ref{defnuapp}) is dependent on
$k^2$ and $\gamma$ while being independent of the charge and mass of the scalar.


\section{Extremal Branes: from near-horizon to boundary of  AdS}
In \S2 we investigated a system of dilaton gravity described by the action eq.(\ref{genact}). 
Since we were interested in the behaviour 
when the dilaton had evolved sufficiently far along a run-away direction we took $f(\phi)$ and $V(\phi)$ to be of the form, eq.(\ref{dildepf}), eq.(\ref{dildeppot}). 
The resulting  solution was then of the form eq. (\ref{ansatz2}). 
In this appendix we will show that 
 such a  solution can arise as the near-horizon limit  starting from an asymptotic $AdS_4$ geometry perturbed by 
a varying dilaton. 
For this purpose we will continue to take $f(\phi)$ to be of the form in eq.(\ref{dildepf}), 
but instead of eq. (\ref{dildeppot}) now
 take the  potential to be 
\be
\label{fullpotapp}
V(\phi)=2 V_0 \cosh(2\delta \phi)
\ee
with $V_0<0$. 
This potential has the property that along the run-away direction where,  $\phi\rightarrow \infty$, 
$V(\phi)\rightarrow V_0 e^{2\delta \phi}$ and therefore   agrees with eq.(\ref{dildeppot}).
As a result the solution eq.(\ref{ansatz2}) continues to be a good approximate solution for this potential as well. 

 In addition, the potential eq.(\ref{fullpotapp})  also has a maximum at $\phi=0$,
with $V(\phi=0)=2 V_0 <0$. This means that   the system has another solution where $\phi=0$ and the metric is  $AdS_4$ with a 
radius  
\be
\label{radsapp}
R^2=-{3\over V_0}.
\ee
Working in a coordinate system of the form eq.(\ref{ansatz1}), 
we will construct a numerical solution which asymptotes   between this $AdS_4$ solution
and a near-horizon geometry given by eq.(\ref{ansatz2}). 

Note that expanding about $\phi=0$ the potential eq.(\ref{fullpotapp}) results in a mass  for the dilaton,
\be
\label{negmapp}
m^2=-2\delta^2 |V_0|.
\ee
In order this to  lie above the BF bound, $m_{BF}^2=-{9\over 4 R^2}$,   $\delta$ must meet the condition 
\be
\label{conddeltapp}
\delta^2<{3\over 8}.
\ee
Asymptotically, as $r\rightarrow \infty$ and one goes towards the boundary of $AdS_4$, the dilaton goes like
\be
\label{falloffd}
\phi\rightarrow r^{\Delta_\pm}, \ \ \Delta_\pm = {-3 \pm\sqrt{9+4m^2}\over 2}
\ee
Since $m^2<0$,  in both cases the dilaton will fall-off. This corresponds to the fact that  with $m^2<0$ the 
 dilaton corresponds to a relevant operator in the CFT dual to the asymptotic 
$AdS_4$ space-time. In the solution we obtain numerically, in general, the dilaton will go like a  linear combination of both 
solutions,
\be
\label{behdil}
\phi=c_1 r^{\Delta_+}+ c_2 e^{\Delta_-}
\ee
Accordingly, in the dual  field theory  the  Lagrangian will be deformed by turning on the relevant operator dual to the dilaton.

In the subsequent discussion it will be convenient to choose units such that  $|V_0|=1$. 

\subsection{Identifying The Perturbation}
It is actually convenient to start in the near-horizon region and then integrate outwards, towards the boundary, to construct the full solution. 

To start,  we first  identify a perturbation in the near-horizon region  which grows  as one goes towards the UV (larger values of $r$).
For this purpose, we will    approximate the potential as $V=- e^{2\delta \phi}$ and ignore the correction going 
like $e^{-2\delta \phi}$ to it,
this will lead to a condition on the parameters $(\alpha,\delta)$ which we will specify shortly. 

Including a perturbation in the metric gives,
\begin{equation}\label{correctionsab}
a(r) = C_a r^{\gamma}(1+ \epsilon \ d_1 \ r^{\nu_1}) \hspace{5mm} ;\hspace{5mm}  b(r)= r^\beta(1+ \epsilon \ d_2 \ r^{\nu_2})
\end{equation}
The resulting form of the perturbation of $\phi$ is determined from the ansatz for $b$ by the equation of motion \eqref{em2}:
\begin{equation}\label{correctionsphi}
 \phi(r)= k \log{r} + \epsilon \ d_3 \ r^{\nu_2}
\end{equation}
with $d_3 = {4 ( \nu_2-1 ) + (\alpha + \delta)^2 (1 + \nu_2) \over 4 (\alpha + \delta)}d_2$. We now proceed to find a solution with $\nu_1 = \nu_2$\footnote{Other choices might give solutions but we have not studied them.}.

Solving  eq.\eqref{em1} and eq.\eqref{em3} to leading order in $r$, determine $\nu_1$ and  
$d_2$ in terms of $d_1$:
\begin{eqnarray}\label{nu1root}
\nu_1&=&-{3\over 2}+{4+2\delta(\alpha+\delta) \over 4 + (\alpha + \delta)^2}\nonumber \\
&&+
{\sqrt{(4+(3\alpha-\delta)(\alpha+\delta))\left[36-(\alpha+\delta)(17\delta-19 \alpha +8 \alpha^2 \delta+8 \alpha \delta^2)\right]} \over 2(4 + (\alpha + \delta)^2)^2}
\end{eqnarray}
and
\begin{equation}\label{d1d2}
d_2={-2 (\alpha + \delta)^2 \over 4 (\nu_1-1 ) + (\alpha + \delta) (\alpha ( \nu_1-1 ) + \
\delta (3 + \nu_1))} d_1.
\end{equation}
Note that $d_1$ is left undetermined and is a   free parameter that characterises the resulting solution. 
It is also worth noting that the perturbation we have identified satisfies the constraint, eq.(\ref{em4}). 

\begin{figure}
\begin{center}
\includegraphics[scale=0.7]{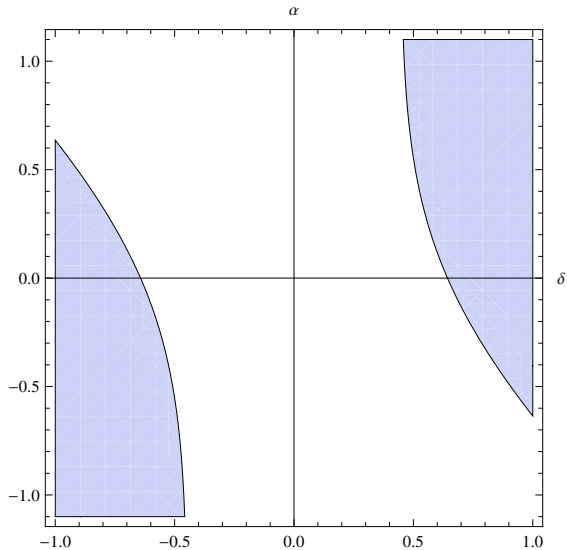}
\end{center}
\caption{\label{plotalbe} Region in $\alpha,\delta$ plane satisfying  $\nu_1 < -4 \delta k$.}
\end{figure}

In our analysis above to determine the perturbation we approximated the potential
$V=-2\cosh(2\delta \phi) \simeq -e^{2\delta \phi}$, while keeping the leading corrections due to the perturbation,
eq.(\ref{correctionsab}), eq.(\ref{correctionsphi}).  This is justified,  for small $r$ if, 
\be
\label{condpertapp}
\nu_1<-4 \delta k.
\ee
Figure \ref{plotalbe}  shows the region in the  $(\alpha,\delta)$ plane which is allowed by this constraint. 
In the numerical analysis we will choose values for $(\alpha, \delta)$ which lie in this region, and 
 which also meet the condition eq.(\ref{conddeltapp}).  

\subsection{Numerical integration}

Staring with the perturbed solution in the near-horizon region the equations can be now be 
numerically integrated to obtain the solution 
for larger values of $r$. For this purpose the full potential eq.(\ref{fullpotapp}) is used.

\begin{figure}
  \begin{center}
  \includegraphics[scale=0.7]{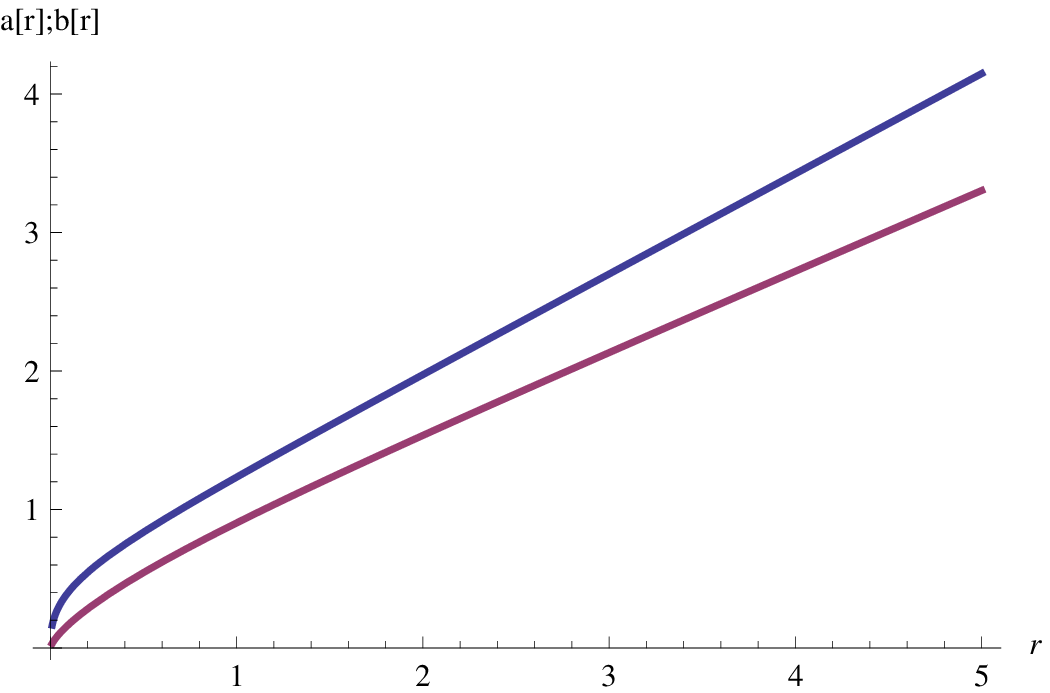}
  \includegraphics[scale=0.7]{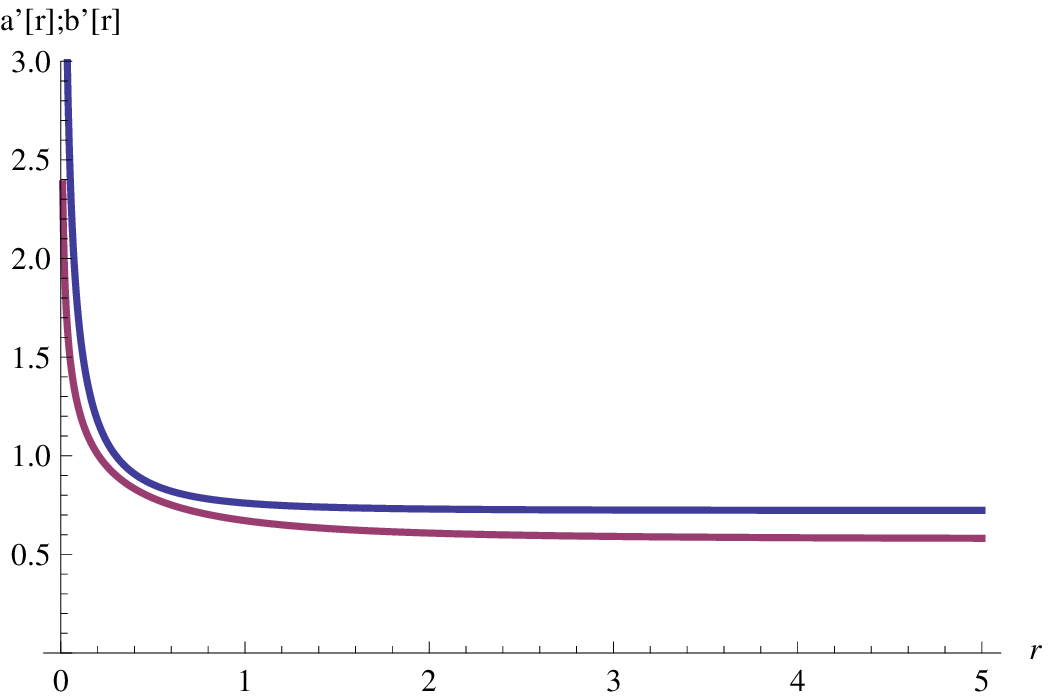}
  \end{center}
\caption{ \label{plotA} Numerical solution interpolating between the near horizon solution and $AdS_4$ for $\alpha=1$, $\delta=0.6$ and $d_1 =0.01$. The second plot shows that $a'(r)$ and $b'(r)$ approach $1$. Red lines denote $a$, Blue lines denote $b$.} 
\end{figure}

\begin{figure}
\begin{center}
 \includegraphics[scale=0.7]{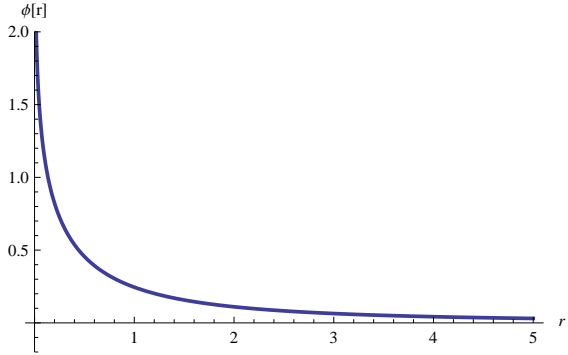}
\end{center}
\caption{\label{plotPhi} Numerical solution for $\phi$, for $\alpha=1$, $\delta=0.6$.}
\end{figure}

Figure \ref{plotA} and Figure \ref{plotPhi} show the resulting solution for $\alpha=1$, $\delta=0.6$, these values satisfy the 
conditions, eq.(\ref{conddeltapp}), eq.(\ref{condpertapp}).  The strength of the perturbation was
chosen to be  $d_1 =0.01$.
 

Figure \ref{plotA} and Figure \ref{plotPhi} clearly show that $a(r) \propto r$ and $b(r) \propto r$ for large $r$, so the solution is
asymptotically $AdS_4$ (A coordinate transformation can be used to set the constant of proportionality to 1 as in the standard $AdS_4$ space). The dilaton approaches $0$, the extrema of the potential $\cosh(2 \delta\phi)$.
Thus the solution interpolates between $AdS_4$  and  solution discussed in \S2 in the near-horizon region. 

Qualitatively similar results are obtained \footnote{ The value,   $\delta =0.6$, is such that both quantisations for the dilaton are possible. Our conclusion that qualitatively similar results are obtained 
continues to also  hold 
when $|\delta|$ is chosen to be somewhat smaller so that only the standard quantisation is allowed.}  
if the parameters $\alpha, \delta $ and $d_1$  are varied within a range \footnote{
We work in the region where eq.(\ref{condpertapp}) and eq.(\ref{conddeltapp}) are met.}. The solution continues to asymptote to $AdS_4$ and the dilaton asymptotes
to $\phi=0$ which is the extremum of $V$. In particular, $\alpha=1, \delta=0.6,$ for which the results 
are presented in Fig \ref{plotA} and Fig \ref{plotPhi}, corresponds to $\beta+\gamma>1$. The range of values
 for which we have found qualitatively similar behaviour includes also cases where $\beta+\gamma\le 1$.

One final comment about parameters. For a given $\alpha$ and $\delta$ there should be a two-parameter family of solutions
 corresponding to the chemical potential $\mu$ and the coupling constant  of the relevant operator dual to the dilaton in the boundary theory. 
Our solution above has one parameter $d_1$ which is the strength of the perturbation in the IR.
Another parameter, which can be thought of as changing   the overall energy scale in the boundary theory, corresponds to 
a coordinate rescaling  in the bulk, $(r, x^\mu)\rightarrow(\lambda r, x^\mu/\lambda)$. Under this coordinate transformation
the charge $Q$, eq.(\ref{gf}), transforms as $Q_e\rightarrow \lambda^2 Q_e$.



\end{document}